\documentclass[reprint,amsmath,amssymb,aps,twocolumn,superscriptaddress]{revtex4-2}

\pdfoutput=1

\usepackage{graphicx}
\usepackage{mathtools}
\usepackage{dcolumn}
\usepackage{bm}
\usepackage{float}
\usepackage[utf8]{inputenc}
\usepackage{booktabs}
\usepackage{tabularx}
\usepackage{colortbl}
\usepackage{etoolbox}
\usepackage{titlesec}
\usepackage{braket}
\usepackage{xcolor}
\usepackage[font=small]{caption}

\titlespacing*{\section}{0pt}{0.5\baselineskip}{0.4\baselineskip}
\titlespacing*{\subsection}{0pt}{0.5\baselineskip}{0.4\baselineskip}

\begin{document}

\title{Poynting vector controversy in axion modified electrodynamics}

\author{Michael E Tobar}
\email{michael.tobar@uwa.edu.au}
\affiliation{Quantum Technologies and Dark Matter Labs, Department of Physics, University of Western Australia, 35 Stirling Highway, Crawley, WA 6009, Australia.}
\author{Ben T McAllister}
\affiliation{Quantum Technologies and Dark Matter Labs, Department of Physics, University of Western Australia, 35 Stirling Highway, Crawley, WA 6009, Australia.}
\author{Maxim Goryachev}
\affiliation{Quantum Technologies and Dark Matter Labs, Department of Physics, University of Western Australia, 35 Stirling Highway, Crawley, WA 6009, Australia.}

\begin{abstract}

The most sensitive haloscopes that search for axion dark matter through the two photon electromagnetic anomaly convert axions into photons through the mixing of axions with a large background direct current (DC) magnetic field. In this work we apply the Poynting theorem to the resulting axion modified electrodynamics and identify two possible Poynting vectors, one which is similar to the Abraham Poynting vector in electrodynamics and the other to the Minkowski Poynting vector. Inherently the conversion of axions to photons is a non-conservative process with respect to the created oscillating photonic degree of freedom. We show that Minkowski Poynting theorem picks up the added nonconservative terms while the Abraham does not. The non-conservative terms may be categorized more generally as ``curl forces", which in classical physics are non-conservative and non-dissipative forces localized in space, not describable by a scalar potential and exist outside the global conservative physical equations of motion. To understand the source of energy conversion and power flow in the detection systems, we apply the two different Poynting theorems to both the resonant cavity haloscope and the broadband low-mass axion haloscope. Our calculations show that both Poynting theorems give the same sensitivity for a resonant cavity axion haloscope, but predict markedly different sensitivity for the low-mass broadband capacitive haloscope. Hence we ask the question, can understanding which one is the relevant one for axion dark matter detection be considered under the framework of the Abraham-Minkowski controversy? In reality, this should be confirmed by experiment when the axion is detected. However, many electrodynamic experiments have ruled in favor of the Minkowski Poynting vector when considering the canonical momentum in dielectric media. In light of this, we show that the axion modified Minkowski Poynting vector should indeed be taken seriously for sensitivity calculation for low-mass axion haloscopes in the quasi static limit, and predict orders of magnitude better sensitivity than the Abraham Poynting vector equivalent.

\end{abstract}


\maketitle

\section{INTRODUCTION}

Axions are postulated to exist as neutral spin-zero bosons to solve the strong charge-parity problem in QCD. Such a particle is predicted to couple very weakly to other known particles and has thus been postulated to be cold dark matter \cite{PQ1977,Wilczek1978,Weinberg1978,wisps,K79,Kim2010,Zhitnitsky:1980tq,DFS81,SVZ80,Dine1983,Preskill1983,Sikivie1983,Sikivie1983b}. In particular, many experiments rely on the electromagnetic anomaly, which is a two photon coupling term with the axion. To gain a significant sensitivity, it is widely considered that the best way to search for the axion is when the first background photonic degree of freedom is a large direct current (DC) magnetic field, which generates a second photon that can be detected. This is the basis of the DC magnetic field axion haloscope, first proposed by Sikivie \cite{Sikivie83haloscope,Sikivie1984} and pioneered experimentally by the ADMX Collaboration \cite{hagmann1990,Hagmann1990b,Bradley2003,ADMXaxions2010,ADMX2011,Braine2020,Bartram2021}. Recently the scientific case that dark matter may include QCD axions or axionlike particles of varied mass and photon coupling has gained momentum \cite{Svrcek_2006,Arvanitaki10,Higaki_2013,Baumann16,Co2020,Co2020b,Co2021,Oikonomou21,Sikivie2021,Sokolov:2021uv,DILUZIO20201,Rodd2021}, leading to many new ideas and new detectors designs worldwide, which implement the principle of dark matter detection through the electromagnetic anomaly \cite{Payez_2015,McAllisterFormFactor,Gupta2016,McAllister:2016fux,XSWisp,ABRACADABRA,ABRA21,McAllister2017,MadMax17,Millar_2017,Ioannisian_2017,Majorovits_2020,PhysRevD.96.123008,JEONG2018412,IRASTORZA201889,Oue19,Nagano2019,Goryachev2019,PhysRevD.96.061102,Henning2019,Liu2019,Marsh2019,Sch_tte_Engel_2021,Lawson2019,Anastassopoulos2017,ZhongHS,Kultask20,TOBAR2020,Gelmini2020,berlin2020axion,Lasenby2020b,Gramolin:2021wm,Abeln:2021us,Thomson2021,universe7070236,10.1093/ptep/ptab051,Devlin2021,Kwon2021,Backes:2021wd,IWAZAKI2021115298,Chigusa:2021vh,PhysRevD.103.096001}. 

Inherently, the conversions of axions among a DC magnetic background field into a second photonic degree of freedom is a nonconservative process. In this process, the axion mixes with the background DC field to create a source term that drives energy conversion to the second photonic degree of freedom, at a frequency corresponding to the axion mass. In this work we implement Poynting theorem with respect to axion modified electrodynamics to understand the energy flow in such a system. 

In standard electrodynamics, Poynting vector analysis is implemented in circuit and antenna theory to understand how the input source power is impressed into the system along with the system power flow and how it relates to the stored energy and losses \cite{White1999,KeqianZhang2008,Volkakis2012,Balanis2012,RHbook2012}. In this work we undertake a similar analysis within the framework of axion modified electrodynamics. In other work, the Poynting vector has been implemented with versions of the stress-energy tensor to understand energy and forces in magnetic and dielectric matter. For example, forces in systems such as optical tweezers \cite{Wu2009,Berry_2013,Bethune_Waddell_2015,Sukhov_2017} and trapping of particles \cite{SpinCurlF2013,Guha:2020uh}, where the best way to analyze these systems has been a subject of controversy (known as the Abraham–Minkowski controversy), is still an active area of debate \cite{Nelson91,Griffiths2012,Mansuripur:2017us,TORCHIGIN2020164986,RevModPhys.79.1197,Leonhardt:2006ux,Barnett2010}. This debate has led to the general concept of ``curl forces," which are abundant in nature, cannot be described from the gradient of a scalar potential, and only exist in a localized space \cite{Wu2009,Berry_2013,SpinCurlF2013,Berry2015}. For example, the force on a particle with complex electric polarizability is known not to be derivable from a scalar potential as its curl is nonzero. Such forces are nonconservative and nondissipative, and their inclusion has been described both classically and quantum mechanically \cite{Strange_2018,Berry_2020}, in particular the quantising of electrodynamics in dielectric and dispersive media \cite{Drezet2016,Drezet2017,Drezet17b}. Note, such nonconservative curl forces do not include the most well-known curl force, which is the magnetic Lorentz force, as it is a conservative force that can do no work \cite{Strange_2018}, described by a magnetic vector potential. 

With this in mind, it has become evident that it is possible to derive alternative versions of Poynting's theorem (in fact, four versions are possible)\cite{Kinsler_2009}. In particular, the Minkowski Poynting vector \cite{Minkowski1908}, $\vec{S}_{D B}=\frac{1}{\epsilon_0\mu_0}\vec{D} \times \vec{B}$, has been shown to be successful to account for experiments in dielectric media, where the field momentum is associated with the canonical momentum \cite{Barnett2010,Chiao2004,Griffiths2012,RevModPhys.79.1197}; here the electric flux density, $\vec{D}=\epsilon_{0} \vec{E}+\vec{P}$, is the sum of the electric field, $\vec{E}$, and electric polarization, $\vec{P}$, and the magnetic flux density, $\vec{B}=\mu_{0}(\vec{H}+\vec{M})$, is the sum of the magnetic field, $\vec{H}$, and magnetization, $\vec{M}$. Naturally, when the curl of the polarization is nonzero ($\nabla\times\vec{P}\ne 0$) the Minkowski Poynting vector will pick up this term, due to an unconventional but necessary modification to Faraday's law \cite{Kinsler_2009,Drezet2016}, while the Abraham Poynting vector \cite{Abraham:2009vq,Abraham:1910wn}, $\vec{S}_{E H}=\vec{E} \times \vec{H}$, will not. 

For the curl of the polarization to be nonzero, an energy input is required to separate the bound charge; this describes a permanent electret or energy harvesting material \cite{Tobar2021} as well as the properties of ferroelectric domain walls \cite{Vasudevan:2017uf}. This description is also similar to an active dipole in antenna theory, a voltage source in circuit theory, or an active dipole emitter in quantum theory \cite{Drezet2016,Drezet2017,Drezet17b}, where an external nonconservative force (sometimes referred as a fictitious or pseudo force) is described by an impressed electric field (some times referred as a fictitious or pseudo electric field) \cite{Pikulin16,Yu19,Ilan:2020to} with a nonzero curl (one could call this a polarization). Furthermore, the electret, energy harvester, or ferroelectric domain may be classified as an active bound charge dipole. We may recognize this active dipole term generally as a nonconservative curl force term, which necessarily modifies Faraday's law, and is only present internally to the active antenna, voltage source, electret, or ferroelectric domain and not present globally outside the active device. As with all curl forces, this nonconservative term cannot be characterized by a scalar potential; on the other hand, it has been recently shown to be characterized via an electric vector potential \cite{Drezet2016,Drezet2017,Tobar2021b,TobarModAx19,TOBAR2020,Tobar2021}. 

Recently, it was also shown that there exists a similar nonconservative curl force term in axion modified electrodynamics  \cite{TobarModAx19,TOBAR2020}. This occurs when the axion mixes with a DC background magnetic field, which converts the axion mass to the energy of the second photonic degree of freedom \cite{TobarModAx19,TOBAR2020}. In this representation the axion mixing with the DC magnetic field adds a similar term to a polarization with a nonzero curl \cite{Tobar2021,Tobar2021b}. In this work we apply the Minkowski and Abraham Poynting theorem equivalents to axion modified electrodynamics and compare the difference, where the former picks up the extra curl force term, while the latter does not.

\section{THE EFFECTIVE AXION CURRENT AND CHARGE DENSITY}

It is well known that axions modify electrodynamics through the axion two photon anomaly \cite{Wilczek:1987aa,Sikivie2021}, which in vacuum becomes:
\begin{equation}
\begin{aligned}
&\nabla \cdot\vec{E}=\frac{\rho_{e}}{\varepsilon_0}+cg_{a \gamma \gamma} \vec{B}.\nabla a \\
&\nabla \times\vec{B}-\frac{1}{c^2} \partial_t\vec{E}=\mu_0 \vec{J}_{e}-g_{a\gamma\gamma}\mu_0\epsilon_0c\left(\vec{B}\partial_t a+\nabla a\times\vec{E}\right)\\
&\nabla\cdot \vec{B}=0\\
&\nabla \times \vec{E}+\partial_t \vec{B}=0.
\end{aligned}
\label{ModAxED}
\end{equation}
Here $g_{a \gamma \gamma}$ is the two-photon coupling to an axion field, $a(t)$ is the amplitude of the axion field, $\rho_e$ is the volume charge density, and $\vec{J}_{e}$ is the volume current density. One common way to set up the equations of motion for the two photon interaction is to assume $\nabla a=0$, so two of the three terms go to zero and only one modification to Ampere's law remains,
\begin{equation}
\nabla \times\vec{B}-\frac{1}{c^2} \partial_t\vec{E}=\mu_0 \left(\vec{J}_{e}-g_{a\gamma\gamma}\epsilon_0c\vec{B}\partial_t a\right),
\end{equation}
where the axion current is defined by
\begin{equation}
\vec{J}_{a}=-g_{a\gamma\gamma}\epsilon_0c\vec{B}\partial_t a.
\label{AXCur}
\end{equation}
This modification is commonly used in the calculation of the sensitivity for haloscope experiments.

A more general version of the modifications as source terms can be obtained by substituting the following vector identities: $\vec{B}\cdot\nabla a=\nabla\cdot(a\vec{B})-a(\nabla\cdot \vec{B})$ and $\nabla a\times\vec{E} =\nabla\times(a\vec{E})-a(\nabla \times \vec{E})$ into (\ref{ModAxED}). Then, assuming to first order $\nabla\cdot\vec{B}=0$ and $\nabla\times\vec{E}=-\partial_t\vec{B}$, the modified Gauss' and Ampere's laws, may be written as \cite{Sikivie2021,TobarModAx19}
\begin{equation}
\begin{aligned}
&\epsilon_0\nabla \cdot\vec{E}=\rho_{e}+\rho_{ab}\\
&\frac{1}{\mu_0}\nabla \times\vec{B}-\epsilon_0\partial_t\vec{E}=\vec{J}_{e}+\vec{J}_{ab}+\vec{J}_{ae},
\end{aligned}
\label{AxEDCurr}
\end{equation}
where
\begin{equation}
\begin{aligned}
&\rho_{ab}=g_{a \gamma \gamma}\epsilon_0c\nabla \cdot\left(a(t)\vec{B}(\vec{r},t)\right)\\
&\vec{J}_{ab} = -g_{a\gamma\gamma}\epsilon_0c\partial_t\left(a(t)\vec{B}(\vec{r},t)\right)\\
&\vec{J}_{ae} = -g_{a\gamma\gamma}\epsilon_0c\nabla\times\left(a(t)\vec{E}(\vec{r},t)\right)
\end{aligned}
\label{AxEDCurr2}
\end{equation}
Here, $\vec{J}_{ab}$ is similar to a polarization current, $\rho_{ab}$ is similar to a bound charge, and they are related through the continuity equation
\begin{equation}
\nabla\cdot\vec{J}_{ab} = - \partial_t \rho_{ab}. \label{eq:C6}
\end{equation}
Furthermore, $\vec{J}_{ae}$ is similar to a bound current, so the total axion current is thus $\vec{J}_a=\vec{J}_{ab}+\vec{J}_{ae}$, which is a more general form of En. (\ref{AXCur}). Note that setting these terms to zero because $\nabla a=0$, is an approximation that is not always valid \cite{TobarModAx19,TOBAR2020}, as we show in the next section.

\section{AXION MODIFIED ELECTRODYNAMICS}

\subsection{Time dependent form}

Rather than write the equation of motion with modified source terms, as in (\ref{AxEDCurr}) and (\ref{AxEDCurr2}), we may include the modifications in the definitions of the fields themselves in a similar way to the auxiliary fields in matter \cite{TobarModAx19}. For example, the macroscopic description should in principal be similar to the way Maxwell's equations are modified due to a sum over a large number of microscopic quantum spins in a permanent magnet. Summing up the effects of all the individual spins leads to macroscopic definition of the magnetization vector, $\vec{M}$, as an auxiliary field. This auxiliary field may also be described by a fictitious bound electric current density at the boundary of the permanent magnet, given by $\vec{J}_b=\nabla\times\vec{M}$. Classically, this ``fictitious" boundary source drives a magnetomotive force (mmf) and creates a magnetic field. We call the current ``fictitious" as there is no real free charge flow, however $\vec{J}_b$ takes the place in Ampere's law in the same place the free charge current would also be. This just means that either a free current loop or permanent magnet may create a magnetic field or mmf, and is the reason we also call an excited wire wound as a coil an ``electromagnet". Applying the same principal to a large ensemble of axions, we show that we may represent the axion-photon anomaly as modifications to Maxwell's equations due to the addition of similar macroscopic auxiliary fields. 

Rearranging the source terms of Eqs. (\ref{AxEDCurr}) and (\ref{AxEDCurr2}) to the left-hand side of Eq.(\ref{ModAxED}) it may be written as \cite{Younggeun18,Tobar2021,Sikivie2021}
\begin{equation}
\begin{aligned}
&\nabla \cdot\left(\vec{E}(\vec{r},t)-g_{a \gamma \gamma} a(t)c\vec{B}(\vec{r},t)\right)=\frac{\rho_{e}}{\epsilon_0}\\
&\nabla \times\left(c\vec{B}(\vec{r},t)+g_{a \gamma \gamma} a(t) \vec{E}(\vec{r},t)\right)\\ 
&-\frac{1}{c} \partial_t\left(\vec{E}(\vec{r},t)-g_{a \gamma \gamma} a(\vec{r},t)c\vec{B}(\vec{r},t)\right)=c\mu_0\vec{J}_{e}\\
&\nabla \cdot c\vec{B}(\vec{r},t)=0\\
&\nabla \times \vec{E}(\vec{r},t)+\frac{1}{c}\partial_t c\vec{B}(\vec{r},t)=0.
\end{aligned}
\label{Reacted}
\end{equation}
This has been shown to be equivalent to a perturbative transformation of the electromagnetic fields \cite{VISINELLI,TobarModAx19,Asker2018}, given by
\begin{equation}
c\vec{B}^{\prime}(\vec{r},t)\rightarrow c\vec{B}(\vec{r},t)+g_{a\gamma\gamma}a(t)\vec{E}(\vec{r},t)~~\text{and}
\label{Bfield}
\end{equation}
\begin{equation}
\vec{E}^{\prime}(\vec{r},t)\rightarrow\vec{E}(\vec{r},t)-g_{a\gamma\gamma}a(t)c\vec{B}(\vec{r},t),
\label{Efield}
\end{equation}
where Eqs. (\ref{Bfield}) and (\ref{Efield}) in the quasistatic limit, effectively represent dual symmetry with respect to a rotation angle, $\theta(t)=g_{a\gamma\gamma}a(t)$ where $\theta(t)\ll 1$ \cite{Cameron_2012,Cameron_2012b,Bliokh_2013,Bliokh_2016}. Here $\theta(t)$ is an effective dynamical pseudoscalar field, which in this case is the product of the axion pseudoscalar field, $a(t)$, with the axion photon coupling, $g_{a\gamma\gamma}$. For dark matter axions, $a(t)$ is in general a large classical field; however, $\theta(t)$ remains small due to the extremely weak coupling of axions to photons, i.e., $g_{a\gamma\gamma}\ll 1$. Note that there is also a duality transformation between electromagnetic potentials, where the dual 4-vector potential contains a magnetic scalar potential and an electric vector potential. Under this duality transform the electric vector potential manifests \cite{Cameron_2012,Cameron_2012b,Bliokh_2013,Bliokh_2016,Asker2018}, which potentially adds the axion induced curl force to the system under investigation. This is evident from Eq. (\ref{Efield}), as the curl of $\vec{E}^{\prime}_1$ has a nonzero spatial term.

Now considering the interaction includes two photons, we distinguish between a background field (denoted by subscript 0) and the generated photon field (denoted by subscript $1$), which is created by the axion pseudoscalar field mixing with the background field. To first order we may assume the background field satisfies Maxwell's equations, so that
\begin{equation}
\begin{aligned}
&\nabla \times \vec{B}_0=\mu_0\epsilon_0\partial_{t} \vec{E}_0+\mu_0\vec{J}_{e_0} \\
&\nabla \times \vec{E}_0=-\partial_{t} \vec{B}_0\\
&\nabla \cdot \vec{B}_0=0 \\
&\nabla \cdot \vec{E}_0=\epsilon_0^{-1}\rho_{e_0}
\end{aligned}
\label{bground}
\end{equation}
Note that any axion modification of the background field will end up second order with respect to the effects on the second generated photonic degree of freedom, so it can be ignored\cite{Younggeun18,Sikivie2021}. 

Thus for the generated photonic degree of freedom, we may write (\ref{Reacted}) in a similar way to how the auxiliary fields are included in Maxwell's equations in matter, and we find axion modified electrodynamics in a familiar form\cite{TobarModAx19}, given by
\begin{equation}
\begin{aligned}
&\nabla \cdot\vec{D}_1=\rho_{e1}\\
&\nabla \times\vec{H}_1-\partial_t\vec{D}_1=\vec{J}_{e1}\\
&\nabla \cdot \vec{B}_1(\vec{r},t)=0\\
&\nabla \times \vec{E}_1(\vec{r},t)+\partial_t\vec{B}_1(\vec{r},t)=0,
\end{aligned}
\label{Da}
\end{equation}
where (\ref{Bfield}) and (\ref{Efield}) are akin to the following constitutive relations:
\begin{equation}
\begin{aligned}
\vec{H}_1(\vec{r},t)&=\frac{\vec{B}_1}{\mu_{0}}-\vec{M}_1-\vec{M}_{1a}~~\text{and} \\
\vec{D}_1(\vec{r},t)&=\epsilon_{0} \vec{E}_1+\vec{P}_1+\vec{P}_{1a}.
\end{aligned}
\label{DHfield}
\end{equation}
Here, $\vec{M}_1$ and $\vec{P}_1$ are the nonaxion induced magnetization and polarization respectively, while the axion modifications, $\vec{M}_{1a}$ and $\vec{P}_{1a}$ are moved to redefinitions of the auxiliary fields rather than source terms and to first order with respect to the background field are given by
\begin{equation}
\begin{aligned}
\vec{M}_{1a}&=-g_{a\gamma\gamma}a(t)c\epsilon_0\vec{E}_0(\vec{r},t)~~\text{and} \\
\frac{1}{\epsilon_0}\vec{P}_{1a}&=-g_{a\gamma\gamma}a(t)c\vec{B}_0(\vec{r},t),
\end{aligned}
\label{Const}
\end{equation}

Here it is clear the divergence of $\vec{M}_{1a}$ is nonzero, similar to what occurs at the boundaries of a permanent magnet, the curl of $\vec{P}_{1a}$ is nonzero similar to what occurs at the boundaries of a permanent electret, and by combining (\ref{Const}) and (\ref{bground}) it can be calculated to be (assuming $\nabla a=0$)
\begin{equation}
\begin{aligned}
\nabla\cdot\vec{M}_{1a}&=-g_{a\gamma\gamma}a(t)c\epsilon_0\nabla\cdot\vec{E}_0(\vec{r},t)=-g_{a\gamma\gamma}a(t)c\rho_{e_0}\\
\frac{1}{\epsilon_0}\nabla\times\vec{P}_{1a}&=-g_{a\gamma\gamma}a(t)c\nabla\times\vec{B}_0(\vec{r},t)\\
&=-\frac{g_{a\gamma\gamma}a(t)}{c}\partial_{t} \vec{E}_0-g_{a\gamma\gamma}a(t)c\mu_0\vec{J}_{e_0}.
\end{aligned}
\label{DivCurl}
\end{equation}
Note that if we followed the procedure to set $\nabla a=0$ at the start of the calculation, then the axion current in (\ref{AXCur}) would be the only modification, and the general form of the modified constitutive relations in (\ref{DHfield}) would be missed. This would be akin to falsely setting $\nabla\times\vec{P}_{1a}=0$ and $\nabla\cdot\vec{M}_{1a}=0$ even though they are in general nonzero in the approximation when $\nabla a$ is set to zero.

Assuming only a DC background magnetic field, $\vec{B}_0(\vec{r})$ with no background electric field ($\vec{E}_0=0$ and $\vec{M}_{1a}=0$) as well as in vacuum ($\vec{M}_{1}=0$, $\vec{P}_1=0$, $\vec{B}_1=\mu_0\vec{H}_1$), one can write the axion modified Ampere's law from Eq. (\ref{Da})  as
\begin{equation}
\nabla \times \vec{B}_1=\mu_0\partial_t \vec{D}_{1}+\mu_0\vec{J}_{e_1}
\label{AmpD}
\end{equation}
In contrast, Faraday's law with respect to the $\vec{E}_1$ and $\vec{B}_1$ fields remains unchanged,
\begin{equation}
\nabla \times \vec{E}_1=-\partial_t\vec{B}_1.
\label{FarE}
\end{equation}
However, given that the curl of $\vec{P}_{1a}$ in Eq. (\ref{DivCurl}) is nonzero ($\frac{1}{\epsilon_0}\nabla\times\vec{P}_{1a}=-g_{a\gamma\gamma}a(t)c\mu_0\vec{J}_{e_0}$ for the DC case), a modified Faraday's law may be written in a similar fashion as to a voltage source (such as a energy harvester, ferroelectret or electricity generator) when the curl is nonzero \cite{Kinsler_2009,Drezet2016,Drezet2017,Tobar2021,Tobar2021}, so by taking the curl of $\vec{D}_{1}$ in (\ref{DHfield}) and combing with (\ref{FarE}) we obtain
\begin{equation}
\frac{1}{\epsilon_0}\nabla \times \vec{D}_{1}=-\partial_t \vec{B}_1-g_{a\gamma\gamma}a\mu_0c\vec{J}_{e_0},
\label{FarD}
\end{equation}
which is analogous to an electromagnetic system in matter where the curl of the polarization is nonzero. It has been shown in such systems the fundamental electromagnetic quantities become the electric $\vec{D}$ and and magnetic $\vec{B}$ flux densities \cite{Kinsler_2009,Drezet2016,Tobar2021,Tobar2021b}, which is compatible with the Minkowski Poynting vector.

In this work we apply these more general equations to low-mass axion haloscopes, which necessarily includes the impressed current of an electromagnet, $\vec{J}_{e_0}$, which excites the background magnetic field $\vec{B}_0(\vec{r})$. Note there also exists a dual symmetry with the source terms in the above Eqs. (\ref{AmpD}) and (\ref{FarD}), where an fictitious magnetic current manifests through the axion interaction, so $\vec{J}^{\prime}_{m_1}(\vec{r},t)\rightarrow g_{a\gamma\gamma}a(t)\mu_0c\vec{J}_{e_0}(\vec{r},t)$. The fact that this impressed magnetic current exists does not necessitate the existence of free magnetic poles, in the same way bound currents and polarization currents do not need the existence of free charge carriers. For example, bound magnetic poles exist in nature as permanent magnets consisting of north and south pole pairs, which can be set in motion, with a net bound magnetic current if one pole is kept stationary as the other rotates. Such a rotating magnet converts the mechanical motion to an electromotive force with nonzero curl (a curl force) \cite{Tobar2021}. This fact has been recognized as early as 1936 \cite{Schelkunoff36}, where Schelkunoff from Bell Labs stated ``It is true that there are no magnetic conductors and no magnetic conduction currents in the same sense as there are electric conductors and electric conduction currents but magnetic convection currents are just as real as electric convection currents, although the former exist only in doublets of oppositely directed currents since magnetic charges themselves are observable only in doublets."

Moreover, most work in the low mass limit assumes as $m_a\rightarrow 0$, the equations of motion (for example Eq.(\ref{Reacted})) do not contribute an observable effect from the axion modified equations of motion. This is because they are compatible with a total derivative and it is assumed all surface terms vanishes to spatial infinity. However, this is not necessarily true as pointed out by others \cite{Cao2017,TobarModAx19,TOBAR2020}. If surface terms exist and do not go to infinity, the equations given by (\ref{Da})-(\ref{FarD}) do not describe the whole system without specifying the boundary conditions. In general, the external degree of freedom that couples to the macroscopic Maxwell's equations to generate an emf may be modelled with a fictitious magnetic current boundary source. Thus, for the Minkowski case, the classical macroscopic description of an ensemble of quantum axion-photon anomalies sum up to give classical equations of motion similar to that of an alternating current (AC) electricity generator \cite{Tobar2021}. On reflection this is not surprising as the conversion of axions into photons may be viewed as the generation of a photonic degree of freedom (or electricity) from an external source, which is essentially a form of electrical power generation.

\subsection{Harmonic phasor form}

For harmonic solutions of the axion-Maxwell equations we write the equations in complex vector-phasor form. For example, we set $\vec{E}_1(\vec{r},t)=\frac{1}{2}\left(\mathbf{E}_1(\vec{r})  e^{-j \omega_1 t}+\mathbf{E}_1^*(\vec{r}) e^{j \omega_1 t}\right)=\operatorname{Re}\left[\mathbf{E}_1(\vec{r}) e^{-j \omega_1 t}\right]$, so we define the vector phasor (bold) and its complex conjugate by $\tilde{\mathbf{E}}_1(\vec{r},t)=\mathbf{E}_1(\vec{r}) e^{-j \omega_1 t}$ and $\tilde{\mathbf{E}}_1^*(\vec{r},t)=\mathbf{E}_1^*(\vec{r}) e^{j \omega_1 t}$, respectively. In contrast, the axion pseudoscalar field, $a(t)$, may be written as $a(t)=\frac{1}{2}\left(\tilde{a} e^{-j \omega_a t}+\tilde{a} ^* e^{j \omega_a t}\right)= \operatorname{Re}\left(\tilde{a} e^{-j \omega_a t}\right)$, and thus, in phasor form, $\tilde{A}=\tilde{a}e^{-j \omega_a t}$ and $\tilde{A}^*=\tilde{a}^*e^{j \omega_a t}$. Thus, the phasor form of the modified Ampere's law in (\ref{AmpD}) becomes
\begin{equation}
\begin{aligned}
\frac{1}{\mu_0}\nabla \times\tilde{\mathbf{B}}_1 &=\tilde{\mathbf{J}}_{e_1}-j\omega_1\epsilon_0\tilde{\mathbf{E}}_{1}+j\omega_ag_{a\gamma\gamma}\epsilon_0c\tilde{A}\vec{B}_0\\
\frac{1}{\mu_0}\nabla \times\tilde{\mathbf{B}}_1^* &=\tilde{\mathbf{J}}_{e_1}^*+j\omega_1\epsilon_0\tilde{\mathbf{E}}_{1}^*-j\omega_ag_{a\gamma\gamma}\epsilon_0c\tilde{A}^*\vec{B}_0,
\end{aligned}
\label{PhasorAmp}
\end{equation}
while the phasor form of Faraday's law in (\ref{FarE}) becomes
\begin{equation}
\begin{aligned}
\nabla \times \tilde{\mathbf{E}}_1 &=j\omega_1\tilde{\mathbf{B}}_1 \\
\nabla \times \tilde{\mathbf{E}}_1^* &=-j\omega_1\tilde{\mathbf{B}}_1^*,
\end{aligned}
\label{PhasorFar}
\end{equation}
and the phasor form of the modified Faraday's law in (\ref{FarD}) becomes
\begin{equation}
\begin{aligned}
\frac{1}{\epsilon_0}\nabla\times\tilde{\mathbf{D}}_{1}&=j\omega_1\tilde{\mathbf{B}}_1-g_{a\gamma\gamma}c\mu_0\tilde{A}\vec{J}_{e_0}\\
\frac{1}{\epsilon_0}\nabla\times\tilde{\mathbf{D}}_{1}^*&=-j\omega_1\tilde{\mathbf{B}}_1^*-g_{a\gamma\gamma}c\mu_0\tilde{A}^*\vec{J}^*_{e_0}
\end{aligned}
\label{PhasorModFar}
\end{equation}
In the following we use these equations to calculate energy and power via the Poynting theorem in a DC magnetic field axion haloscope.

\section{CALCULATION OF POWER GENERATED IN A DC MAGNETIC FIELD AXION HALOSCOPE USING Poynting THEOREM}

We start by considering the instantaneous Poynting vector in its standard physics textbook form of,
\begin{equation}
\begin{aligned}
\vec{S}_1(t) &=\frac{1}{\mu_0}\vec{E}_1(t) \times \vec{B}_1(t)= \\
&\frac{1}{2}\left(\mathbf{E}_{1}e^{-j \omega_1 t}+\mathbf{E}_{1}^{*}e^{j \omega_1 t}\right) \times \frac{1}{2\mu_0}\left(\mathbf{B}_{1} e^{-j \omega_1 t}+\mathbf{B}_{1}^{*}e^{j \omega_1 t}\right)\\
&=\frac{1}{2\mu_0} \operatorname{Re}\left(\mathbf{E}_{1} \times \mathbf{B}_{1}^{*}\right)+\frac{1}{2\mu_0} \operatorname{Re}\left(\mathbf{E}_{1} \times \mathbf{B}_{1}~e^{-j2\omega_1 t}\right),
\end{aligned}
\label{Inst}
\end{equation}
which consists of a DC term, the first term on the right-hand side of (\ref{Inst}), and a high frequency term, the second term on the right-hand side of (\ref{Inst}). Note, the DC term in (\ref{Inst}) is equivalent to the time average of the instantaneous Poynting vector. 

Thus, the complex Poynting vector and its complex conjugate are defined by,
\begin{equation}
\mathbf{S}_1=\frac{1}{2\mu_0} \mathbf{E}_1 \times \mathbf{B}_1^{*}~~\text{and}~~\mathbf{S}_1^{*}=\frac{1}{2\mu_0} \mathbf{E}_1^{*} \times \mathbf{B}_1,
\label{AbPv}
\end{equation}
respectively, where $\mathbf{S}_1$ is the complex power density of the harmonic electromagnetic wave or oscillation, with the real part equal to the time averaged power density and the imaginary term equal to the reactive power, which may be inductive (magnetic energy dominates) or capacitive (electrical energy dominates). 

In the case that the complex Poynting vector is only real, then the $ \mathbf{E}_1$ and $ \mathbf{B}_1$ fields are in phase, which describes a propagating wave with distinct direction and momentum. Such a propagating wave can be generated by an antenna in the far-field limit, at distances larger than the wavelength of the emitted photon: and is a source of loss from the antenna. Another case where $ \mathbf{E}_1$ and $ \mathbf{B}_1$ are in phase is due to resistive losses: in this case the photon energy is converted to heat and destroyed; however both are effectively loss terms with respect to the antenna, the former known as radiation loss. Conversely, in the near field limit of an antenna, the Poynting vector is imaginary as $ \mathbf{E}_1$ and $ \mathbf{B}_1$ are out of phase. This represents reactive energy flow between the antenna power source and the antenna near field, which exists at subwavelength distances from the antenna. In this case the photons do not propagate away from the antenna, and they exist as quasistatic oscillating $ \mathbf{E}_1$ and $ \mathbf{B}_1$ fields with no net momentum.

A convenient and unambiguous way to calculate the real and imaginary part of the Poynting vector is through the following equations:
\begin{equation}
\begin{aligned}
\operatorname{Re}\left(\mathbf{S}_1\right)=\frac{1}{2}(\mathbf{S}_{1}+\mathbf{S}_{1}^*)~\text{and}~j\operatorname{Im}\left(\mathbf{S}_1\right)=\frac{1}{2}(\mathbf{S}_{1}-\mathbf{S}_{1}^*).
\end{aligned}
\label{ReIm}
\end{equation}
We use these equations in the following to calculate the real and imaginary parts of the complex Poynting vector in axion modified electrodynamics.

\subsection{Axion modified Minkowski Poynting theorem}

Based on the axion modified $\vec{D}_{1}$ vector, we may calculate the complex axion modified Minkowski Poynting vector in a similar way to Eq. (\ref{AbPv}), which is given by,
\begin{equation}
\begin{aligned}
\mathbf{S}_{DB}=\frac{1}{2\epsilon_0\mu_0} \mathbf{D}_{1} \times \mathbf{B}_1^{*}~~\text{and}~~\mathbf{S}_{DB}^{*}=\frac{1}{2\epsilon_0\mu_0} \mathbf{D}_{1}^{*} \times \mathbf{B}_1
\end{aligned}
\label{Mink}
\end{equation}
Taking the divergence of Eq. (\ref{Mink}) we find
\begin{equation}
\begin{aligned}
&\nabla \cdot\mathbf{S}_{DB}=\frac{1}{2} \nabla \cdot(\frac{1}{\epsilon_0}\mathbf{D}_{1} \times \frac{1}{\mu_0}\mathbf{B}_1^*) =\\
&\frac{1}{2}\left(\frac{1}{\mu_0}\mathbf{B}_1^* \cdot\frac{1}{\epsilon_0}\nabla \times \mathbf{D}_{1}-\frac{1}{\epsilon_0}\mathbf{D}_{1} \cdot\frac{1}{\mu_0}\nabla \times \mathbf{B}_1^* \right),
\end{aligned}
\label{Mink2}
\end{equation}
and
\begin{equation}
\begin{aligned}
&\nabla \cdot\mathbf{S}_{DB}^*=\frac{1}{2} \nabla \cdot(\frac{1}{\epsilon_0}\mathbf{D}_{1}^* \times \frac{1}{\mu_0}\mathbf{B}_1) =\\
&\frac{1}{2}\left(\frac{1}{\mu_0}\mathbf{B}_1 \cdot\frac{1}{\epsilon_0}\nabla \times \mathbf{D}_{1}^*-\frac{1}{\epsilon_0}\mathbf{D}_{1}^* \cdot\frac{1}{\mu_0}\nabla \times \mathbf{B}_1 \right).
\end{aligned}
\label{Mink3}
\end{equation}
Combining (\ref{Mink2}) and  (\ref{Mink3}) with (\ref{PhasorModFar}),  (\ref{PhasorAmp}) and (\ref{ReIm}) along with the divergence theorem (this is a standard technique in microwave engineering and circuit theory \cite{Montgomery87,Dicke1987,Volkakis2012,Kinsler_2009}), after some calculation we obtain (see Appendix A for details),
\begin{equation}
\begin{aligned}
&\oint\operatorname{Re}\left(\mathbf{S}_{DB}\right)\cdot \hat{n}ds=\\
&\int\Big(\frac{j(\omega_1-\omega_a)}{4}\epsilon_0g_{a\gamma\gamma}c\vec{B}_0\cdot(\tilde{a}\mathbf{E}_{1}^*-\tilde{a}^*\mathbf{E}_{1})\\
&+\frac{1}{4}g_{a\gamma\gamma}c\vec{B}_0\cdot(\tilde{a}\mathbf{J}_{e_1}^*+\tilde{a}^*\mathbf{J}_{e_1})-\frac{1}{4}g_{a\gamma\gamma}\vec{J}_{e_0}\cdot(\tilde{a}^*c\mathbf{B}_1+\tilde{a}c\mathbf{B}_1^*)\\
&-\frac{1}{4}(\mathbf{E}_1\cdot\mathbf{J}_{e_1}^*+\mathbf{E}_1^*\cdot\mathbf{J}_{e_1})\Big)~dV,
\end{aligned}
\label{MinkRe}
\end{equation}
and
\begin{equation}
\begin{aligned}
&\oint j\operatorname{Im}\left(\mathbf{S}_{DB}\right)\cdot \hat{n}ds=\int\Big(\frac{j\omega_1}{2}\big(\frac{1}{\mu_0}\mathbf{B}_1^* \cdot\mathbf{B}_1-\epsilon_0\mathbf{E}_1\cdot\mathbf{E}_1^*\big)\\
&~+\frac{j(\omega_1+\omega_a)\epsilon_0g_{a\gamma\gamma}}{4}c\vec{B}_0\cdot(\tilde{a}\mathbf{E}_{1}^*+\tilde{a}^*\mathbf{E}_1)~+\\
&\frac{1}{4}g_{a\gamma\gamma}c\vec{B}_0\cdot(\tilde{a}\mathbf{J}_{e_1}^*-\tilde{a}^*\mathbf{J}_{e_1})+\frac{1}{4}g_{a\gamma\gamma}\vec{J}_{e_0}\cdot(\tilde{a}^*c\mathbf{B}_1-\tilde{a}c\mathbf{B}_1^*)\\
&-\frac{1}{4}(\mathbf{E}_1\cdot\mathbf{J}_{e_1}^*-\mathbf{E}_1^*\cdot\mathbf{J}_{e_1})-\frac{j\omega_{a} }{2} \epsilon_{0} g_{a \gamma \gamma}^2a_0^2 c^2 B_{0}^2\Big)~dV
\end{aligned}
\label{MinkIm}
\end{equation}
The closed surface integral on the left-hand side of (\ref{MinkRe}) is the time averaged power radiated from inside to outside the haloscope volume, which for a closed system such as a cavity will be zero. However, for an open system such as a radio frequency antenna, real power will radiate in the far field. In contrast, the closed surface integral on the left-hand side of (\ref{MinkIm}) is the reactive power radiated outside the haloscope volume, which in general does not have to be zero, in a similar way to how reactive power oscillates to and from the voltage source charging and discharging a reactive capacitor in circuit theory, or the reactive power in an antenna, where energy oscillates from the antenna to the near field and then is reabsorbed by the antenna, due to the antenna's self-capacitance or inductance. 

\subsection{Axion modified Abraham Poynting theorem}

The complex Abraham pointing vector is basically the same as Eq. (\ref{AbPv}) for the case we are considering with $\vec{H}_{1}=\frac{1}{\mu_0}\vec{B}_1$. Taking the divergence of Eq. (\ref{AbPv}) we find
\begin{equation}
\begin{aligned}
&\nabla \cdot\mathbf{S}_{EH}=\frac{1}{2} \nabla \cdot(\mathbf{E}_1 \times \mathbf{H}_{1}^*) =\\
&\frac{1}{2}\mathbf{H}_{1}^* \cdot(\nabla \times \mathbf{E}_1)-\frac{1}{2}\mathbf{E}_1 \cdot(\nabla \times \mathbf{H}_{1}^*)
\end{aligned}
\label{Ab2}
\end{equation}
and
\begin{equation}
\begin{aligned}
&\nabla \cdot\mathbf{S}_{EH}^*=\frac{1}{2} \nabla \cdot(\mathbf{E}_1^* \times \mathbf{H}_{1}) =\\
&\frac{1}{2}\mathbf{H}_{1} \cdot(\nabla \times \mathbf{E}_1^*)-\frac{1}{2}\mathbf{E}_1^* \cdot(\nabla \times \mathbf{H}_{1})
\end{aligned}
\label{Ab3}
\end{equation}
Combining (\ref{Ab2}) and  (\ref{Ab3}) with (\ref{PhasorFar}),  (\ref{PhasorAmp}) and (\ref{ReIm}) along with the divergence theorem, we obtain (see Appendix A for details),
\begin{equation}
\begin{aligned}
\oint\operatorname{Re}\left(\mathbf{S}_{EH}\right)\cdot \hat{n}ds&=\int\Big(\frac{j\omega_a}{4}\epsilon_0g_{a\gamma\gamma}c\vec{B}_0\cdot(\tilde{a}^*\mathbf{E}_{1}-\tilde{a}\mathbf{E}_{1}^*))\\
&-\frac{1}{4}(\mathbf{E}_1\cdot\mathbf{J}_{e_1}^*+\mathbf{E}_1^*\cdot\mathbf{J}_{e_1})\Big)~dV
\end{aligned}
\label{AbRe}
\end{equation}
and
\begin{equation}
\begin{aligned}
\oint j\operatorname{Im}\left(\mathbf{S}_{EH}\right)\cdot \hat{n}ds&=\int\Big(\frac{j\omega_1}{2}\big(\mu_0\mathbf{H}_{1}^* \cdot\mathbf{H}_{1}-\epsilon_0\mathbf{E}_1\cdot\mathbf{E}_1^*\big)\\
&+\frac{j\omega_a}{4}\epsilon_0g_{a\gamma\gamma}c\vec{B}_0\cdot(\tilde{a}^*\mathbf{E}_{1}+\tilde{a}\mathbf{E}_{1}^*)\\
&-\frac{1}{4}(\mathbf{E}_1\cdot\mathbf{J}_{e_1}^*-\mathbf{E}_1^*\cdot\mathbf{J}_{e_1}))\Big)~dV
\end{aligned}
\label{AbIm}
\end{equation}
As before, the closed surface integral on the left-hand side of (\ref{AbRe}) and (\ref{AbIm}) is the time averaged power and reactive power radiated outside the haloscope volume, respectively. However, the Abraham Poynting vector misses three extra terms the Minkowski Poynting vector picks up, due to the inclusion of the nonconservative and nondissipative source term described by Eq. (\ref{PhasorModFar}). 

\subsection{Abraham or Minkowski Poynting theorem in axion modified electrodynamics?}

Currently most calculations of haloscope detection sensitivity $\it{a~priori}$ assume the Abraham Poynting vector is valid, with the exception of one or two \cite{TobarModAx19,TOBAR2020}. However, as shown in the Minkowski-Abraham debate over the past century, this is not ``clear-cut" and in a stationary dielectric media where the canonical momentum is under consideration, the Minkowski form agrees with experimental results \cite{Barnett2010,Griffiths2012,RevModPhys.79.1197}. This may be true in axion modified electrodynamics, as we can identify a similar guilty term due to Eq. (\ref{DivCurl}) and hence (\ref{FarD}). The nonzero curl should do active work without adding dissipation and should not be ignored in calculations of axion detector sensitivity. This extra term has all the properties of a curl force \cite{Wu2009,Berry_2013,Berry2015}, which adds spatial terms to the Poynting theorem equations. In the following sections we compare the two ways of determining the power of photonic conversion for various axion haloscope topologies.

\section{RESONANT CAVITY HALOSCOPE}

\begin{figure}[t!]
\includegraphics[width=0.49\textwidth]{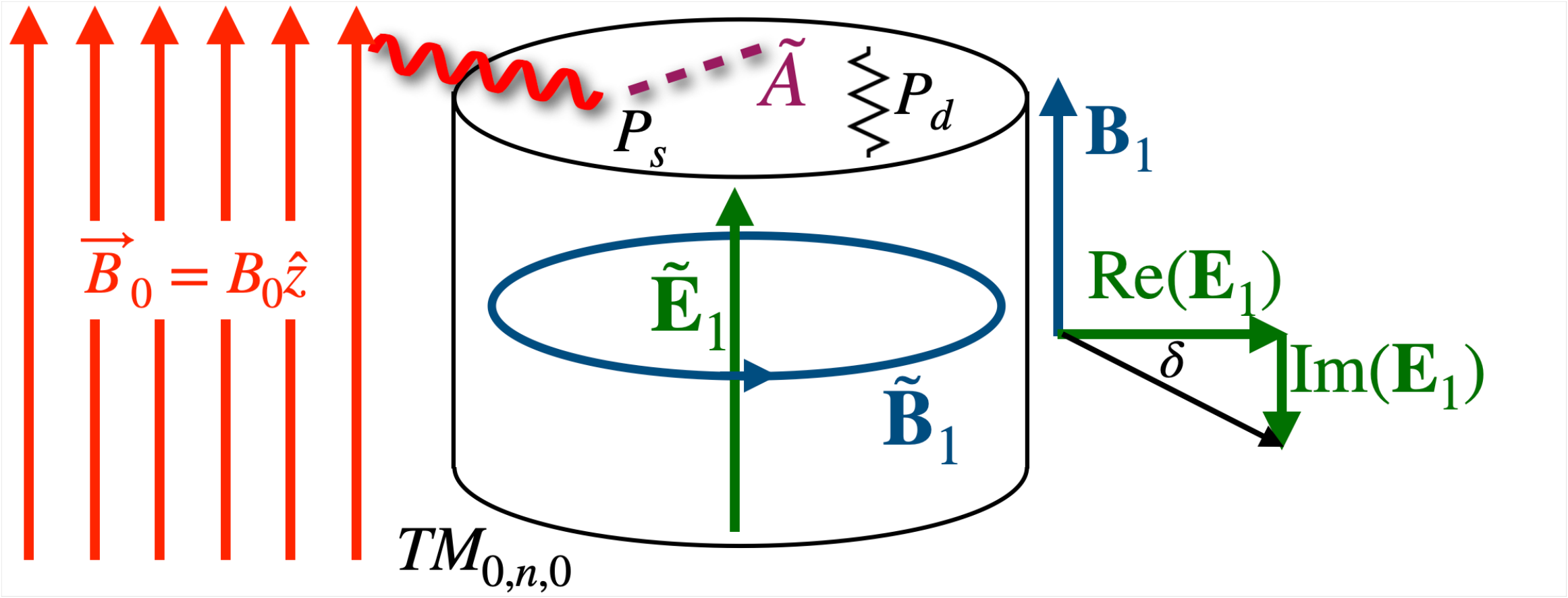}
\caption{Schematic of axion conversion in an ADMX style haloscope. Such haloscopes are based on radio frequency cavity resonators, where an external DC magnetic field, $\vec{B}_0$, has a nonzero dot product with the nondissipative electric field of the radio frequency mode in the cavity, $\operatorname{Re}(\mathbf{E}_1)$. For a cylindrical cavity the sensitive modes are from the $TM_{0,n,0}$ mode family, where $n$ is the radial mode number. The axial and azimuthal mode numbers must be zero. Finite losses mean part of the electric field, $\operatorname{Im}(\mathbf{E}_1)$, is in phase with the magnetic field, $\mathbf{B}_1$, characterized by the loss angle $\delta$, which for a high-$Q$ system is very small and related by $\tan\delta\sim\frac{1}{Q_1}$.}
\label{CavHalo}
\end{figure}

In this section we derive the sensitivity of an ADMX style radio frequency haloscope based on a cavity resonator \cite{Sikivie83haloscope,Sikivie1984,hagmann1990,Hagmann1990b,Bradley2003,ADMXaxions2010,ADMX2011,Braine2020,Bartram2021}, with a schematic shown in Fig. \ref{CavHalo}. First, we undertake the calculation using the Abraham Poynting vector, as this is the $\it{a~priori}$ Poynting vector assumed across most of the literature, and then we compare and contrast calculations using the Minkowski Poynting vector. 

For a power source, $P_s$, delivering energy to a resonator as shown in Fig. \ref{CavHalo}, the resonance is defined when the reactive power delivered by the source is zero, and thus when tuned on resonance the circulating energy only oscillates between the electric and the magnetic energy in the resonator at the cavity resonance frequency, with no energy oscillating between the cavity and the power source (which in this case is the axion mixing with a DC magnetic field). In this case the power delivered to the cavity is real. This corresponds to the real part of the Poynting theorem equations, which we use in the next section to calculate the sensitivity of the axion haloscope. Internal to the cavity resonator, this circulating energy is described by a reactive (or imaginary) Poynting vector, which causes the power in the resonator to build up, with respect to the source input power, $P_{s}$. This buildup is limited by the dissipation in the resonator and hence $Q$-factor. The buildup of circulating power is given by $P_c=Q_1P_s $, where in the steady state $P_{s}=P_d$, which is also related to the stored energy in the resonator, $U_1$, by, $P_{s}=\omega_1U_1$. 

Thus, in such a cavity resonator the electric and magnetic field are out of phase (as opposed to a propagating wave, which is in phase), and in this paper we represent the lossless electric field vector phasor as real and the lossless magnetic field vector phasor as imaginary (and so the cross product is imaginary). Dissipative terms, whether calculated in the volume or on the surface assume Ohm's law, dictate that the dissipative part of the electric field be in phase with surface or volume currents and hence the magnetic field. Thus, the electric field effectively gains an imaginary component when losses are included. However, the majority of the electric field is real, with $\operatorname{Im}(\mathbf{E}_1)\sim-\operatorname{Re}(\mathbf{E}_1)/Q_1$. The tangential real part of the electric field must be continuous at the cavity wall boundary, which for a perfect conductor is zero and sets the boundary conditions to calculate the frequency of the electromagnetic modes. Setting the reactive power to zero on resonance, allows us to calculate if there is any frequency shift of the bare cavity when excited by axions. We do find a small second order in $Q$-factor effect, calculated using Foster's reactance theorem \cite{Foster1924}. However, there is no major impact on the sensitivity calculation. For completeness this is detailed in Appendix B and predicts a different value of frequency shift depending on whether we use the Minkowski or the Abraham Poynting theorem.

\subsection{Cavity dissipated power}

Both Poynting theorems have a dissipative term in the real components, listed as the final term on the right-hand side of Eqs. (\ref{MinkRe}) and (\ref{AbRe}) and graphically shown in Fig. \ref{CavHalo}. For dissipation effects over the volume, the volume current is in phase with the imaginary part of the electric field and is of the form $\mathbf{J}_{e_1}=\sigma_e\mathbf{E}_1$, where $\sigma_e$ is the effective conductivity of the volume, which is related to the loss tangent of the volume by $\sigma_e=\omega_a\epsilon_0\tan\delta$ and given $\frac{1}{Q_1}\sim\tan\delta$, then $\mathbf{J}_{e_1}\sim\frac{\omega_a\epsilon_0}{Q_1}\mathbf{E}_1$, substituting these values in the last term on the right-hand side of Eqs. (\ref{MinkRe}) or (\ref{AbRe}), the dissipated power in the cavity is calculated to be
\begin{equation}
\begin{aligned}
P_d=\frac{\omega_a\epsilon_0}{2Q_1}\int\mathbf{E}_1\cdot\mathbf{E}_{1}^*~dV=\frac{\omega_aU_1}{Q_1}.
\end{aligned}
\label{Pdis}
\end{equation}

For surface loss, the same volume integrals given by (\ref{MinkRe}) and (\ref{AbRe}) collapse to surface integrals, where the surface current on the cavity walls is represented by the vector phasor, $\mathbf{K}_1=\hat{n}\times\mathbf{H}_{1}$, of dimensions amps/meter and the electric field at the surface is nonzero and in phase with the surface current, related by  $\mathbf{E}_{1}=R_S\mathbf{K}_1$, where $R_S$ is the surface resistance. Again, substituting these values in the last term on the right-hand side of Eqs. (\ref{MinkRe}) or (\ref{AbRe}) means the dissipated power in the cavity is
\begin{equation}
\begin{aligned}
P_d=\frac{R_S}{2}\oint\mathbf{K}_1\cdot\mathbf{K}_{1}^*ds=\frac{R_S}{2}\oint\mathbf{H}_{1}\cdot\mathbf{H}_{1}^*ds=\frac{\omega_aU_1}{Q_1},
\end{aligned}
\label{PdisSurf}
\end{equation}
which gives the same relationship with respect to the stored energy, $U_1$, and the dissipated power, whether it dissipates over the volume or over the surface of the cavity resonator. 

\subsection{Sensitivity from the Abraham Poynting theorem}

In this calculation we assume the cavity is a closed system so there is no real power radiating outside the cavity volume, and this means the closed surface integral on the left-hand side of (\ref{AbRe}) should be set to zero ($\oint\operatorname{Re}\left(\mathbf{S}_{EH}\right)\cdot \hat{n}ds=0$). In practice power is taken outside the cavity due to the coupling, which in effect loads the cavity $Q$-factor, and this phenomena may be added after the calculation using standard techniques. We also assume that the axion and the resonator frequency coincide ($\omega_1=\omega_a$), and therefore the magnetic and electric energy inside the resonator will be equal, where again the effects of detuning may be added using standard techniques. Under these assumptions Eqs. (\ref{AbRe}) becomes
\begin{equation}
\begin{aligned}
P_s=\frac{jg_{a\gamma\gamma}\omega_ac}{4}\int \vec{B}_0\cdot(\tilde{a}^*\epsilon_0\mathbf{E}_{1}-\tilde{a}\epsilon_0\mathbf{E}_{1}^*)~dV=P_d=\frac{\omega_aU_1}{Q_1}
\end{aligned}
\label{AbReCav}
\end{equation}
Here, $P_s$ is the axion source power and must be real, note the source power is equal to the dissipated power, and as calculated in the last section can occur over the volume and/or over the cavity surface.

For the source power to be nonzero either $\mathbf{E}_{1}$ or the axion scalar field, $\tilde{a}$, has to be imaginary. Since the axion scalar field is assumed to be lossless, we consider only the former to be imaginary, as has been suggested previously \cite{Younggeun18}. The general complex electric field is of the form $\mathbf{E}_{1}\approx(1-j\tan\delta)\operatorname{Re}(\mathbf{E}_{1})$ in the regime where the loss angle is very small, $\delta\ll 1$. Hence, the axion source term in the steady state becomes
\begin{equation}
\begin{aligned}
P_s=\frac{g_{a\gamma\gamma}a_0\omega_a\epsilon_0c}{2Q_1}\int \vec{B}_0\cdot\operatorname{Re}(\mathbf{E}_{1}(\vec{r}))~dV,
\end{aligned}
\label{Ps}
\end{equation}
where $a_0=\frac{1}{2}(\tilde{a}+\tilde{a}^*)$ is the peak value of the scalar axion field, so $a_0=\sqrt{2}\langle a_0\rangle$. Equating (\ref{Ps}) to $P_d= \frac{\omega_aU_1}{Q_1}$ derived in (\ref{Pdis}) or (\ref{PdisSurf}) gives
\begin{equation}
\begin{aligned}
U_1=\frac{g_{a\gamma\gamma}a_0\epsilon_0c}{2}\int \vec{B}_0\cdot\operatorname{Re}(\mathbf{E}_{1})~dV=\frac{\epsilon_0}{2}\int\mathbf{E}_1\cdot\mathbf{E}_{1}^*~dV.
\end{aligned}
\label{U1}
\end{equation}
Now defining the form factor of the cavity haloscope as 
\begin{equation}
\begin{aligned}
C_1=\frac{\left(\int \vec{B}_0\cdot\operatorname{Re}(\mathbf{E}_{1})~dV\right)^2}{B_0^2V_1\int\mathbf{E}_1\cdot\mathbf{E}_{1}^*~dV},
\end{aligned}
\label{FormFac}
\end{equation}
the axion induced circulating power may be calculated to be
\begin{equation}
\begin{aligned}
P_{1}=&\omega_aQ_1U_1=g_{a\gamma\gamma}^2\langle a_0\rangle^2\omega_aQ_1\epsilon_0c^2B_0^2V_1C_1 \\
=&g_{a\gamma\gamma}^2\rho_{a}Q_1\epsilon_0c^5B_0^2V_1C_1\frac{1}{\omega_a},
\end{aligned}
\label{FormFac2}
\end{equation}
where $\langle a_{0}\rangle^2=\frac{\rho_{a}}{c} \frac{\hbar^2}{m_{a}^2}$ and $\rho_{a}$ is the axion dark matter density. This calculation is consistent with what has been derived previously \cite{Sikivie1984,Sikivie2021,McAllisterFormFactor}.

\subsection{Sensitivity from the Minkowski Poynting theorem}

As before, assuming the real power inside the cavity haloscope is a closed system ($\oint\operatorname{Re}\left(\mathbf{S}_{DB}\right)\cdot \hat{n}ds=0$), the cavity is embedded inside a magnet ($\vec{J}_{e_0}\cdot(\tilde{a}c\mathbf{B}_1^*+\tilde{a}^*c\mathbf{B}_1)=0$), and the axion and the resonator frequency coincide ($\omega_1=\omega_a$). Then, in this case Eq. (\ref{MinkRe}) becomes, 
\begin{equation}
\begin{aligned}
&P_s=\int\frac{1}{4}g_{a\gamma\gamma}c\vec{B}_0\cdot(\tilde{a}\mathbf{J}_{e_1}^*+\tilde{a}^*\mathbf{J}_{e_1})~dV =P_d=\frac{\omega_aU_1}{Q_1}
\end{aligned}
\label{MinkReCav}
\end{equation}
Here, $P_s$ is the axion source power and must be real.

As undertaken in the Abraham technique, we assume a lossy volume current in phase with the electric field of the form $\mathbf{J}_{e_1}=\sigma_e\mathbf{E}_1$ where, $\sigma_e=\frac{\omega_a\epsilon_0}{Q_1}$. Substituting the same value of $\mathbf{J}_{e_1}$ into (\ref{MinkReCav}) gives,
\begin{equation}
\begin{aligned}
P_s=\frac{g_{a\gamma\gamma}a_0\omega_a\epsilon_0c}{2Q_1}\int \vec{B}_0\cdot\operatorname{Re}(\mathbf{E}_{1}(\vec{r}))~dV,
\end{aligned}
\label{PdisM}
\end{equation}
the same as calculated for the Abraham technique in Eq. (\ref{Ps}), which means both Eqs. (\ref{FormFac}) and (\ref{FormFac2}) are calculable using both the Minkowski and the Abraham Poynting vectors, and are consistent with previous sensitivity calculations for a standard ADMX style haloscope.

\section{LOW-MASS BROADBAND AXION HALOSCOPES UNDER DC MAGNETIC FIELD}

For a low-mass broadband detector in the quasistatic limit, a haloscope may be inductive or capacitive and must be driven by reactive power from the source, so in the first approximation any dissipation or radiation loss can be ignored, and is thus set to zero. As before, we consider the generated electric field to be real ($\mathbf{E}_1^*=\mathbf{E}_1$) and the out of phase magnetic field as imaginary, ($\mathbf{B}_1^*=-\mathbf{B}_1$). Also, conduction currents will be in the same phase as the magnetic field and hence imaginary ($\mathbf{J}_{e_1}^*=-\mathbf{J}_{e_1}$). In this case, it is clear that the real part of the delivered complex Poynting vector given by (\ref{AbRe}) and (\ref{MinkRe}) must be zero, and the sensitivity of the reactive low-mass broadband haloscope will be determined from the imaginary reactive power delivered by the axion interacting with the background DC magnetic field. 

There has been some recent controversy in the calculation of sensitivity for low mass reactive experiments in the quasi static limit, where the majority of the publications suggest that the sensitivity to electric field is suppressed when the Compton wavelength of the axion is larger than the experimental dimensions \cite{Ouellet2018,Younggeun18,Beutter_2019,ABRACADABRA}. These experiments assume that the only modification to Maxwell's equations is due to the axion current (\ref{AXCur}), which is equivalent to assuming no boundary or spatial effects and thus setting the total derivative to zero. On the other hand, it has been shown that making these approximations too early in the calculation can lead to valid solutions being lost \cite{Cao2017,TobarModAx19,TOBAR2020} due to extra spatial or surface terms, which in this case is due to the fact that $\vec{P}_{1a}=-g_{a\gamma\gamma}a(t)\epsilon_0c\vec{B}_0(\vec{r})$ has a nonzero curl, which can also be thought as a connection to the Witten effect\cite{Witten:1979ua,rodrgueztzompantzi2020}. Based on this, more sensitive experiments have been proposed using inductive wire loop readouts \cite{TOBAR2020}, or capacitive parallel plate readouts \cite{BEAST,TobarModAx19}. In the following, as an example, we compare the sensitivity of a parallel plate capacitor to low mass axions by implementing both Poynting vector theorems.

\subsection{Capacitor under DC magnetic field}

\begin{figure}[t!]
\includegraphics[width=0.45\textwidth]{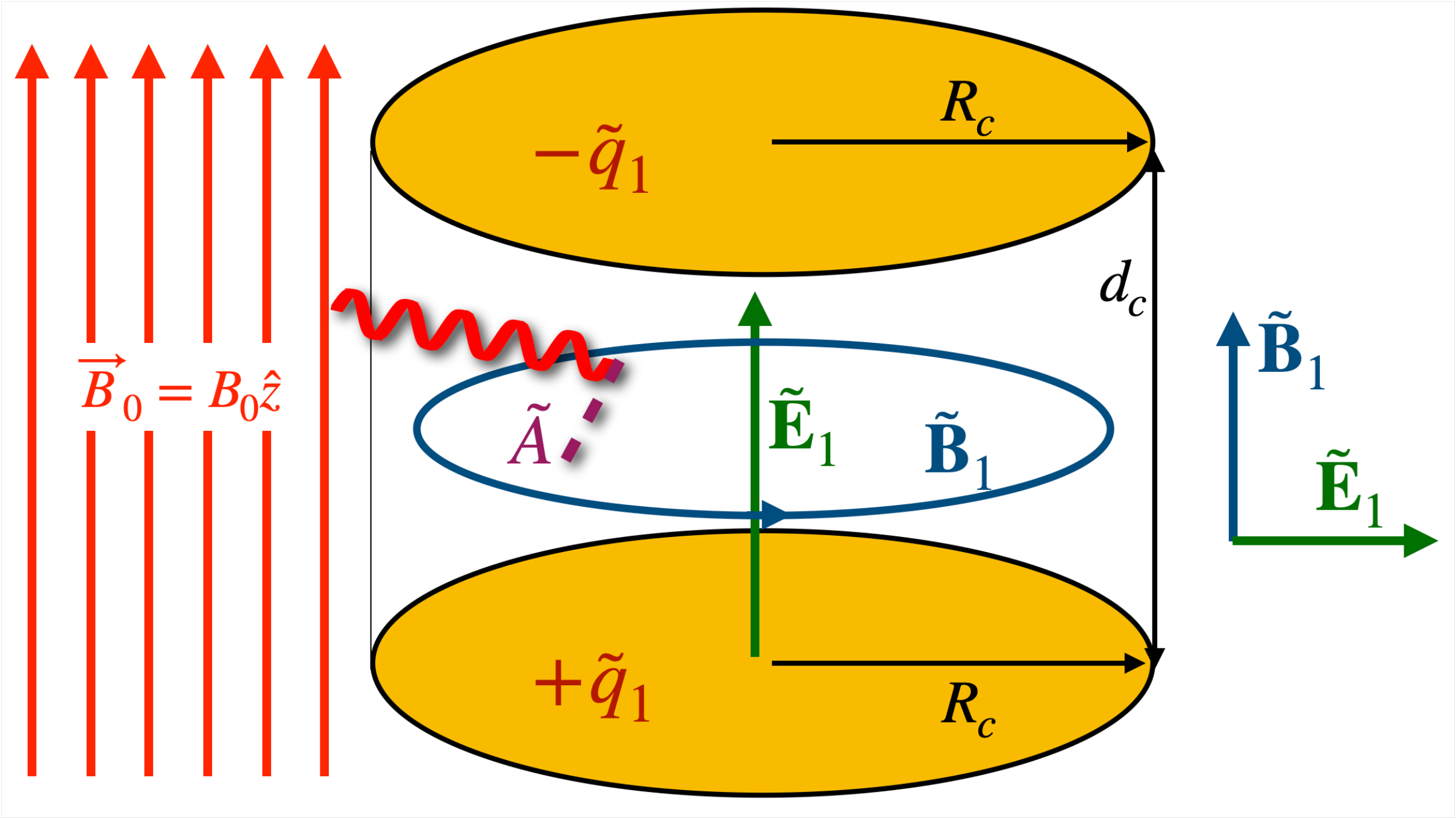}
\caption{Schematic of axion conversion in a capacitive haloscope}
\label{HaloCap}
\end{figure}

For a parallel plate capacitor as shown in Fig.\ref{HaloCap} the last terms on the right-hand side of Eqs. (\ref{AbIm}) and (\ref{MinkIm}) must be zero, since the conduction current must be zero in the lossless capacitor volume. This also means the third last term on the right-hand side of (\ref{MinkIm}) must be zero. Furthermore, if we assume the capacitor is embedded inside a DC magnet, the second last term in (\ref{MinkIm}) must also be zero (it is possible to make use of this term to make a sensitive low-mass detector \cite{TOBAR2020}). This means the equations for reactive power flowing into and out of the capacitor volume, using the Abraham and Minkowski forms, are given by
\begin{equation}
\begin{aligned}
&\oint j\operatorname{Im}\left(\mathbf{S}_{EH}\right)\cdot \hat{n}ds=j\omega_a\int\Big(\frac{1}{2\mu_0}\mathbf{B}_1^* \cdot\mathbf{B}_1-\frac{\epsilon_0}{2}\mathbf{E}_1^*\cdot\mathbf{E}_1\\
&+\frac{\epsilon_0}{2}g_{a\gamma\gamma}a_0c\vec{B}_0\cdot Re(\mathbf{E}_{1})\Big)~dV
\end{aligned}
\label{AbImCap}
\end{equation}
and
\begin{equation}
\begin{aligned}
&\oint j\operatorname{Im}\left(\mathbf{S}_{DB}\right)\cdot \hat{n}ds=j\omega_a\int\Big(\frac{1}{2\mu_0}\mathbf{B}_1^*\cdot\mathbf{B}_1-\frac{\epsilon_0}{2}\mathbf{E}_1^*\cdot\mathbf{E}_1\\
&+\epsilon_0g_{a\gamma\gamma}a_0c\vec{B}_0\cdot Re(\mathbf{E}_{1})-\frac{1 }{2} \epsilon_{0} g_{a \gamma \gamma}^2a_0^2 c^2B_0^2 \Big)~dV,
\end{aligned}
\label{MinkImCap}
\end{equation}

For the capacitor in Fig.\ref{HaloCap} the AC electric field phasor, ignoring fringing is of the form
\begin{equation}
\begin{aligned}
\mathbf{E}_1=\frac{\tilde{q}_1}{\pi R_c^2\epsilon_0}\hat{z}=\frac{\tilde{\sigma}_1}{\epsilon_0}\hat{z},
\end{aligned}
\label{ECap}
\end{equation}
where $\tilde{q}_1$ is the complex phasor of electric charge on the capacitor plates and $\tilde{\sigma}_1$ the effective surface charge density. Following this, from Ampere's law the AC magnetic field phasor within the capacitor volume ($V_c=\pi R_c^2d_c$) may be calculated to be
\begin{equation}
\begin{aligned}
\mathbf{B}_1=-j\omega_a\mu_0\tilde{q}_1\frac{r}{2\pi R_c^2}\hat{\theta},
\end{aligned}
\label{BCap}
\end{equation}
Following this we may calculate the ratio of the magnetic energy density to electric energy density in the capacitor given by
\begin{equation}
\begin{aligned}
\frac{\mathbf{B}_1\cdot\mathbf{B}_1^*}{\epsilon_0\mu_0\mathbf{E}_1\cdot\mathbf{E}_1^*}=\frac{r^2\omega_a^2}{4c^2}=\frac{\pi^2r^2}{\lambda_a^2},
\end{aligned}
\label{EBratDens}
\end{equation}
where $\lambda_a$ is the Compton wavelength of the axion. Integrating over the volume of the capacitor, allows us to calculate the ratio of magnetic to electric energy to be (ignoring fringing)
\begin{equation}
\begin{aligned}
\frac{\int_{V_c}\mathbf{B}_1\cdot\mathbf{B}_1^*dV}{\epsilon_0\mu_0\int_{V_c}\mathbf{E}_1\cdot\mathbf{E}_1^*dV}=\frac{R_c^2\omega_a^2}{8c^2}=\frac{\pi^2R_c^2}{2\lambda_a^2}.
\end{aligned}
\label{EBrat}
\end{equation}
These equations highlight that at DC the parallel plate capacitor is purely capacitive, but at AC the capacitor has a small but finite inductance in the quasi static limit, when $\lambda_a>R_c$. When $\lambda_a\sim R_c$, the capacitor could become resonant, similar to a $TM_{0,1,0}$ mode in a cylindrical cavity; however, this would not be in the quasistatic limit. Nevertheless, in a circuit where the direction of the electric field $\mathbf{E}_{1}$ in a capacitor is parallel to the applied DC magnetic field, $\vec{B_0}$, Eq.(\ref{U1}) still holds for the capacitor, with an effective form factor of unity, which can be shown by substituting Eq. (\ref{ECap}) into (\ref{FormFac}), and we can use this fact to help calculate the sensitivity of a low-mass capacitor experiment.

\subsubsection{Sensitivity assuming the Abraham Poynting theorem}

Assuming the Abraham Poynting theorem, the reactive power delivered to and from a capacitor under the DC magnetic field as shown in Fig.\ref{HaloCap} can be calculated by substituting Eq. (\ref{U1}) into (\ref{AbImCap}), and using (\ref{EBratDens}) we find
\begin{equation}
\begin{aligned}
&jP_{a}=\oint j\operatorname{Im}\left(\mathbf{S}_{EH}\right)\cdot \hat{n}ds=\\
&\frac{j\omega_ag_{a\gamma\gamma}a_0\epsilon_0c}{2}\int\big(\vec{B}_0\cdot \operatorname{Re}(\mathbf{E}_{1})\big)\frac{\pi^2r^2}{\lambda_a^2}~dV,
\end{aligned}
\label{AbImCap2}
\end{equation}
Now, from the definition of the haloscope form factor (\ref{FormFac}) and setting it to be equal to unity, the reactive power delivered to the capacitor given by (\ref{AbImCap2}) becomes (with some algebra)
\begin{equation}
\begin{aligned}
P_a=\omega_aU_c,~ \text{where}~U_c=g_{a\gamma\gamma}^2\langle a_0\rangle^2\epsilon_0c^2B_0^2V_1\Big(\frac{\pi^2Rc^2}{2\lambda_a^2}\Big)^2
\end{aligned}
\label{ReacPow}
\end{equation}
Thus, the magnitude of the voltage phasor across the capacitor can be calculated from  $U_c=\frac{1}{2}\tilde{\mathcal{V}} \tilde{\mathcal{V}}^{*}C_{a}$ ($C_a=\frac{\pi R_c^2\epsilon_0}{d_c}$) to be
\begin{equation}
\lvert\tilde{\mathcal{V}}\rvert=\sqrt{2}~g_{a\gamma\gamma} \langle a_0\rangle c B_0 d_c \Big(\frac{\pi R_c}{\sqrt{2}\lambda_a}\Big)^2
\end{equation}
which means the magnitude of the electric field vector phasor created in the capacitor is,
\begin{equation}
\lvert\mathbf{E}_1\rvert=\sqrt{2}~g_{a\gamma\gamma} \langle a_0\rangle c B_0 \Big(\frac{\pi R_c}{\sqrt{2}\lambda_a}\Big)^2,
\label{EVP}
\end{equation}
agrees with other calculations \cite{Ouellet2018,Younggeun18,Beutter_2019}. Thus, from (\ref{ECap}) the magnitude of the charge per unit area phasor on the capacitor plates will be,
\begin{equation}
\lvert\tilde{\sigma}_1\rvert=\epsilon_0\lvert\mathbf{E}_1\rvert=\sqrt{2}~g_{a\gamma\gamma} \langle a_0\rangle \epsilon_0c B_0 \Big(\frac{\pi R_c}{\sqrt{2}\lambda_a}\Big)^2,
\end{equation}
and the magnitude of the magnetic flux density phasor between the capacitor plates will be,
\begin{equation}
\lvert\mathbf{B}_1\rvert=~\frac{g_{a\gamma\gamma} \langle a_0\rangle\omega_aB_0r}{\sqrt{2}c} \Big(\frac{\pi R_c}{\sqrt{2}\lambda_a}\Big)^2,
\label{MVP}
\end{equation}
and the rms voltage across the capacitor will be,
\begin{equation}
\begin{aligned}
\mathcal{V}_{rms}=&g_{a\gamma\gamma} \langle a_0\rangle c B_0 d_c\Big(\frac{\pi R_c}{\sqrt{2}\lambda_a}\Big)^2 \\
 =&g_{a \gamma \gamma} d_c\frac{c}{\omega_{a}} B_{0} \sqrt{\rho_{a} c^{3}}\Big(\frac{\pi R_c}{\sqrt{2}\lambda_a}\Big)^2,
\end{aligned}
\end{equation}
where $\langle a_{0}\rangle=\sqrt{\frac{\rho_{a}}{c}} \frac{\hbar}{m_{a}}$ and $\rho_{a}$ is the axion dark matter density.
This calculation is consistent with other calculations based on just the axion current \cite{Ouellet2018,Younggeun18,Beutter_2019}, as given by Eq.(\ref{AXCur}): However , it does not take into account the nonzero value of the curl of $\vec{P}_{1a}$. The calculation predicts suppressed sensitivity at low mass, proportional to $\frac{R_c^2}{\lambda_a^2}$. Note, in this scenario the reactive power is positive and excited through the capacitors non-zero inductance.

\subsubsection{Sensitivity assuming the Minkowski Poynting theorem}

Assuming the Minkowski Poynting theorem, the reactive power delivered to and from a capacitor under the DC magnetic field as shown in Fig.\ref{HaloCap} can be calculated by substituting the electric and magnetic field vector phasor given by Eq. (\ref{EVP}) and (\ref{MVP}) into (\ref{MinkImCap}). For this case only the last term in (\ref{MinkImCap}) is significant, and we obtain
\begin{equation}
\begin{aligned}
&jP_{a}=\oint j\operatorname{Im}\left(\mathbf{S}_{DB}\right)\cdot \hat{n}ds\approx \\
&-\frac{j\omega_a}{2}\int\big(\epsilon_{0} g_{a \gamma \gamma}^2a_0^2 c^2B_0^2\big)~dV,
\end{aligned}
\label{MinkImCap2}
\end{equation}
Integrating over the volume the energy stored in the capacitor becomes
\begin{equation}
\begin{aligned}
U_c=g_{a\gamma\gamma}^2\langle a_0\rangle^2\epsilon_0c^2B_0^2V_1.
\end{aligned}
\label{ReacPow}
\end{equation}
Thus, the magnitude of the voltage phasor across the capacitor can be calculated from  $U_c=\frac{1}{2}\tilde{\mathcal{V}} \tilde{\mathcal{V}}^{*}C_{a}$ to be,
\begin{equation}
\lvert\tilde{\mathcal{V}}\rvert=\sqrt{2}~g_{a\gamma\gamma} \langle a_0\rangle c B_0 d_c,
\end{equation}
which is consistent with an $rms$ voltage of,
\begin{equation}
\mathcal{V}_{rms}=g_{a\gamma\gamma} \langle a_0\rangle c B_0 d_c =g_{a \gamma \gamma} d_c\frac{c}{\omega_{a}} B_{0} \sqrt{\rho_{a} c^{3}},
\end{equation}
which is the same as calculated previously \cite{BEAST}. Thus, we may conclude, from the Minkowski Poynting theorem, a sensitive low-mass experiment may be undertaken using a capacitive haloscope. Note, in this scenario the reactive power is negative and excited through the capacitance, this excitation will then create a non-zero secondary electric field across the capacitor, which is not suppressed, this is different to (\ref{EVP}), which is created directly by the axion interaction.

\section{DISCUSSION AND CONCLUSIONS}

By applying the Poynting theorem to axion modified electrodynamics, we have shown how the sensitivity of a resonant cavity and reactive broadband axion haloscope may be calculated. However, the way we apply the theorem is dependent on the type of detector. For example, the Poynting vector analysis has already been undertaken to calculate the sensitivity of the MADMAX detector \cite{MadMax17,Millar_2017,Ioannisian_2017,Majorovits_2020}. However, MADMAX is in the regime where the Compton wavelength of the axion is much smaller than the detector size, and it is thus in a different regime from the resonant and reactive haloscope discussed in this paper. The MADMAX detector converts energy at a dielectric boundary and is assumed to be in the propagating wave (or far field) limit, where the $\mathbf{\tilde{E}}$ and $\mathbf{\tilde{B}}$ vector phasors are in phase, so the Poynting vector is real, represents the physical energy flux leaving a surface, propagates through the haloscope \cite{MadMax17,Millar_2017,Ioannisian_2017}, and in principle can be made broadband.

In contrast, the resonant haloscope is generally the size of the Compton wavelength of the axion (unless higher order modes are implemented) and thus has an imaginary Poynting vector internally within the resonator. This is because the axion induced photon energy produced within the resonator is reflected at the resonator boundaries, so the energy is localized in the form of a standing wave, with the $\mathbf{\tilde{E}}$ and $\mathbf{\tilde{B}}$ fields out of phase. In this work we have assumed the electric field is real, and thus the magnetic field is imaginary. However, on resonance (when $\omega_a=\omega_1$), the axion conversion process within the resonant cavity haloscope does not need to supply any reactive power, only real power. In this case the real part of the Poynting theorem equation has both a source term and a dissipative term within the cavity, which are equal in the steady state, allowing the incident source power to escape the volume as heat, through the resistive losses. Meanwhile, reactive power flow oscillates between the electric and the magnetic fields within the cavity. The higher the $Q$-factor the more the circulating power builds up within the cavity, meaning the percentage of dissipation per cycle is smaller, and hence the detector sensitivity is proportional to the $Q$-factor. The down side is that the technique is narrow band, which requires complicated tuning mechanisms to scan for the axion of unknown mass.

On the other hand, low-mass broadband experiments are in the quasistatic regime, where the Compton wavelength is much greater than the dimensions of the detector. In this case the sensitivity is determined by the reactive power flow within the detector created from the axion-photon conversion. For the higher frequency resonant cavity haloscope, we have shown that the implementation of either the Minkowski or the Abraham axion modified Poynting vector has no significant influence on the calculated sensitivity. In contrast, for low-mass reactive haloscopes there is a large difference in sensitivity calculated from the two Poynting theorems. 

Currently, the Minkowski–Abraham controversy in electrodynamics interacting with matter is considered to be resolved by identifying the Abraham and Minkowski Poynting vector with the total system kinetic and canonical momentum respectively \cite{Barnett2010,Barnett2010b}. For electrodynamics in matter, the Abraham Poynting theorem is the correct one to use when the whole dielectric body is displaced together as a solid entity \cite{Griffiths2012}. In contrast, the Minkowski Poynting vector is the relevant one to use when considering the canonical momentum, which acts to spatially translate particles within the dielectric \cite{Chiao2004}, such as bound charge, which may in some cases cause the curl of the polarization to be nonzero \cite{Kinsler_2009,Tobar2021,Vasudevan:2017uf,Drezet2016,Drezet2017,Drezet17b}. For example, the conservation law for the canonical momentum has been validated through atomic recoil in spontaneous emission \cite{Chiao2004}.

For axion modified electrodynamics the Minkowski solution suggests a similar nonconservative effect under a DC background magnetic field, $\vec{B}_0(\vec{r})$, because from Eq.(\ref{DivCurl}), when $\nabla a=0$, the curl of $\vec{P}_{1a}$ is nonzero and is given by $\nabla\times\vec{P}_{1a}=-g_{a\gamma\gamma}a(t)c\epsilon_0\nabla\times\vec{B}_0(\vec{r})=-g_{a\gamma\gamma}a(t)\epsilon_0\mu_0c\vec{J}_{e_0}$. This term is created at the surface defined by the coil winding where the impressed electric current, $\vec{J}_{e_0}$, flows, which is the source term that creates $\vec{B}_0(\vec{r})$. The curl of $\vec{P}_{1a}$ may also be described as a fictitious oscillating magnetic current boundary source ($\vec{J}_{m_1}=g_{a\gamma\gamma}a(t)\mu_0c\vec{J}_{e_0}$), which sources an oscillating spatial nonconservative electric curl force (or emf), with a force per unit charge of $-g_{a\gamma\gamma}a(t)c\vec{B}_0$, which displaces the axion charge harmonically in time, creating a polarization current of $\vec{J}_a=\partial_t\vec{P}_{1a}=-g_{a\gamma\gamma}\epsilon_0c\vec{B}_0\partial_t a$, which is equivalent to the axion current given by Eq.(\ref{AXCur}). Here, the harmonic spatial translation of axion charge creates the oscillating photonic degree of freedom through the QCD axion-photon anomaly in an analogous way to photons created via spontaneous emission from dipole emitters in \cite{Chiao2004}. The nonconservative boundary source gives an unsuppressed sensitivity at
low mass proportional to $g_{a\gamma\gamma}a(t)$, whereas experiments proportional to the axion current are suppressed through the time derivative.

Recently, this effect has also been shown to be apparent in topological insulators, where a material with a polarization of nonzero curl was shown to be associated with a magnetic current boundary (or instanton), a Berry phase and nonzero crystal momentum \cite{Song2021}. Moreover, they showed a nonzero static $\theta$ angle angle was possible because a nonzero magnetoelectric angle in 3D, does not obstruct the gauge invariance of polarization density, and $\theta$ can be interpreted as the magnetoelectric polarizability, i.e., a magnetic field induces an extra polarization density, $\Delta\vec{P}$ equivalent to Eq.(\ref{Const}) with a static value of $\theta$ \cite{Song2021,Qi2008,Baldomir21}.

For axion modified electrodynamics the Abraham solution is consistent with the total derivative equal to zero, which is the prevailing view among the axion dark matter community. For the total derivative to be zero, it is well known that all surface terms must go to zero as the Compton wavelength approaches infinity (surfaces are essentially assumed to go to infinity). In this case the axion current is the only source term, which contributes to observable effects. In this work we have challenged this view through the Abraham-Minkowski controversy, while putting forward the idea that the conversion from the axion scalar field to power in the oscillating photonic degree of freedom is just another way to generate photonic power (or electricity) from an external nonphotonic degree of freedom. In this case the underlying microscopic mechanism is the axion-photon anomaly from QCD, where the surrounding ensemble of axions from dark matter mix with the virtual photons from the DC magnetic field to supply the external energy that will generate power in the oscillating photonic degree of freedom.

In the end, to verify which description gives the correct solution will require experimentation, which will only be possible when the axion is discovered.

This work was funded by the ARC Centre of Excellence for Engineered Quantum Systems, CE170100009, and the ARC Centre of Excellence for Dark Matter Particle Physics, CE200100008. BM was also funded by the Forrest Research Foundation.

\section*{APPENDIX A: DERIVATION OF Poynting THEOREM EQUATIONS}

In this appendix we derive Eqs. (\ref{MinkRe}), (\ref{MinkIm}), (\ref{AbRe}) and (\ref{AbIm}) in the main text.

\subsection{Axion modified Minkowski Poynting theorem}

To derive Eq. (\ref{MinkRe}) and (\ref{MinkIm}), we begin with writing the divergence of the real and imaginary parts of $\mathbf{S}_{DB}$ as,

\begin{equation}
\begin{aligned}
\nabla\cdot\operatorname{Re}\left(\mathbf{S}_{DB}\right)=\frac{1}{2}\nabla\cdot(\mathbf{S}_{DB}+\mathbf{S}_{DB}^*)\\
\nabla\cdot\operatorname{Im}\left(\mathbf{S}_{DB}\right)=\frac{1}{2}\nabla\cdot(\mathbf{S}_{DB}-\mathbf{S}_{DB}^*)
\end{aligned}
\label{ReImDiv}
\end{equation}

The next step is to calculate $\nabla\cdot\mathbf{S}_{DB}$ and $\nabla\cdot\mathbf{S}_{DB}^*$,
\begin{equation}
\begin{aligned}
&\nabla \cdot\mathbf{S}_{DB}=\frac{1}{2} \nabla \cdot(\frac{1}{\epsilon_0}\mathbf{D}_{1} \times \frac{1}{\mu_0}\mathbf{B}_1^*) =\\
&\frac{1}{2}\left(\frac{1}{\mu_0}\mathbf{B}_1^* \cdot\frac{1}{\epsilon_0}\nabla \times \mathbf{D}_{1}-\frac{1}{\epsilon_0}\mathbf{D}_{1} \cdot\frac{1}{\mu_0}\nabla \times \mathbf{B}_1^* \right),
\end{aligned}
\label{DivSDB}
\end{equation}
and,
\begin{equation}
\begin{aligned}
&\nabla \cdot\mathbf{S}_{DB}^*=\frac{1}{2} \nabla \cdot(\frac{1}{\epsilon_0}\mathbf{D}_{1}^* \times \frac{1}{\mu_0}\mathbf{B}_1) =\\
&\frac{1}{2}\left(\frac{1}{\mu_0}\mathbf{B}_1 \cdot\frac{1}{\epsilon_0}\nabla \times \mathbf{D}_{1}^*-\frac{1}{\epsilon_0}\mathbf{D}_{1}^* \cdot\frac{1}{\mu_0}\nabla \times \mathbf{B}_1 \right).
\end{aligned}
\label{DivSDBIm}
\end{equation}
In harmonic form, the axion modified Ampere's and Faraday's laws under a background DC $\it{B}$-$\it{field}$ of $\vec{B}_0(\vec{r})$, created by an impressed electrical DC current in the magnet coil, $\vec{J}_{e_0}$, may be written as
\begin{equation}
\begin{aligned}
&\frac{1}{\mu_0}\nabla \times\mathbf{B}_1 =\mathbf{J}_{e_1}-j\omega_1\epsilon_0\mathbf{E}_{1}+j\omega_ag_{a\gamma\gamma}\epsilon_0c\vec{B}_0\tilde{a}\\
&\frac{1}{\mu_0}\nabla \times\mathbf{B}_1^* =\mathbf{J}_{e_1}^*+j\omega_1\epsilon_0\mathbf{E}_{1}^*-j\omega_ag_{a\gamma\gamma}\epsilon_0c\vec{B}_0\tilde{a}^*\\
&\frac{1}{\epsilon_0}\nabla\times\mathbf{D}_{1}=j\omega_1\mathbf{B}_1-g_{a\gamma\gamma}c\mu_0\tilde{a}\vec{J}_{e_0}\\
&\frac{1}{\epsilon_0}\nabla\times\mathbf{D}_{1}^*=-j\omega_1\mathbf{B}_1^*-g_{a\gamma\gamma}c\mu_0\tilde{a}^*\vec{J}_{e_0}.
\end{aligned}
\label{AmpFaraday}
\end{equation}
Substituting Eq. (\ref{AmpFaraday})  into Eq. (\ref{DivSDB}) and (\ref{DivSDBIm}) leads to
\begin{equation}
\begin{aligned}
&\nabla \cdot\mathbf{S}_{DB}=\frac{1}{2\mu_0}\mathbf{B}_1^* \cdot(j\omega_1\mathbf{B}_1-g_{a\gamma\gamma}\tilde{a}c\mu_0\vec{J}_{e_0})-\\
&\frac{1}{2}(\mathbf{E}_1-g_{a\gamma\gamma}\tilde{a}c\vec{B}_0) \cdot(\mathbf{J}_{e_1}^*+j\omega_1\epsilon_0\mathbf{E}_{1}^*-j\omega_ag_{a\gamma\gamma}\epsilon_0\tilde{a}^*c\vec{B}_0)\\
&=\frac{j\omega_1}{2}\left(\frac{1}{\mu_0}\mathbf{B}_1^* \cdot\mathbf{B}_1-\epsilon_0\mathbf{E}_1\cdot\mathbf{E}_1^*\right)+\frac{j\omega_1}{2}\epsilon_0g_{a\gamma\gamma}\tilde{a}c\vec{B}_0\cdot\mathbf{E}_{1}^*\\
&+\frac{j\omega_a}{2}g_{a\gamma\gamma}\epsilon_0\tilde{a}^*c\vec{B}_0\cdot\mathbf{E}_1-\frac{1}{2}\mathbf{E}_1\cdot\mathbf{J}_{e_1}^*+\frac{1}{2}g_{a\gamma\gamma}\tilde{a}c\vec{B}_0\cdot\mathbf{J}_{e_1}^*\\
&-\frac{1}{2}g_{a\gamma\gamma}\tilde{a}c\mathbf{B}_1^* \cdot\vec{J}_{e_0}
\end{aligned}
\label{divReEx}
\end{equation}
and
\begin{equation}
\begin{aligned}
&\nabla \cdot\mathbf{S}_{DB}^*=\frac{1}{2\mu_0}\mathbf{B}_1 \cdot(-j\omega_1\mathbf{B}_1^*-g_{a\gamma\gamma}\tilde{a}^*c\mu_0\vec{J}_{e_0})-\\
&\frac{1}{2}(\mathbf{E}_1^*-g_{a\gamma\gamma}\tilde{a}^*c\vec{B}_0) \cdot(\mathbf{J}_{e_1}-j\omega_1\epsilon_0\mathbf{E}_{1}+j\omega_ag_{a\gamma\gamma}\epsilon_0\tilde{a}c\vec{B}_0)\\
&=\frac{j\omega_1}{2}\left(\epsilon_0\mathbf{E}_1\cdot\mathbf{E}_1^*-\frac{1}{\mu_0}\mathbf{B}_1^* \cdot\mathbf{B}_1\right)-\frac{j\omega_1}{2}\epsilon_0g_{a\gamma\gamma}\tilde{a}^*c\vec{B}_0\cdot\mathbf{E}_{1}\\
&-\frac{j\omega_a}{2}g_{a\gamma\gamma}\epsilon_0\tilde{a}c\vec{B}_0\cdot\mathbf{E}_1^*-\frac{1}{2}\mathbf{E}_1^*\cdot\mathbf{J}_{e_1}+\frac{1}{2}g_{a\gamma\gamma}\tilde{a}^*c\vec{B}_0\cdot\mathbf{J}_{e_1}\\
&-\frac{1}{2}g_{a\gamma\gamma}\tilde{a}^*c\mathbf{B}_1 \cdot\vec{J}_{e_0}.
\end{aligned}
\label{divImEx}
\end{equation}
Now by substituting (\ref{divReEx}) and (\ref{divImEx}) into (\ref{ReImDiv}) we obtain
\begin{equation}
\begin{aligned}
&\nabla\cdot\operatorname{Re}\left(\mathbf{S}_{DB}\right)=\frac{j(\omega_1-\omega_a)}{4}\epsilon_0g_{a\gamma\gamma}c\vec{B}_0\cdot(\tilde{a}\mathbf{E}_{1}^*-\tilde{a}^*\mathbf{E}_{1})\\
&+\frac{1}{4}g_{a\gamma\gamma}c\vec{B}_0\cdot(\tilde{a}\mathbf{J}_{e_1}^*+\tilde{a}^*\mathbf{J}_{e_1})-\frac{1}{4}g_{a\gamma\gamma}\vec{J}_{e_0}\cdot(\tilde{a}^*c\mathbf{B}_1+\tilde{a}c\mathbf{B}_1^*)\\
&-\frac{1}{4}(\mathbf{E}_1\cdot\mathbf{J}_{e_1}^*+\mathbf{E}_1^*\cdot\mathbf{J}_{e_1})
\end{aligned}
\end{equation}
and
\begin{equation}
\begin{aligned}
&\nabla\cdot j\operatorname{Im}\left(\mathbf{S}_{DB}\right)=\frac{j\omega_1}{2}\big(\frac{1}{\mu_0}\mathbf{B}_1^* \cdot\mathbf{B}_1-\epsilon_0\mathbf{E}_1\cdot\mathbf{E}_1^*\big)\\
&~+\frac{j(\omega_1+\omega_a)\epsilon_0g_{a\gamma\gamma}}{4}c\vec{B}_0\cdot(\tilde{a}\mathbf{E}_{1}^*+\tilde{a}^*\mathbf{E}_1)~+\\
&\frac{1}{4}g_{a\gamma\gamma}c\vec{B}_0\cdot(\tilde{a}\mathbf{J}_{e_1}^*-\tilde{a}^*\mathbf{J}_{e_1})+\frac{1}{4}g_{a\gamma\gamma}\vec{J}_{e_0}\cdot(\tilde{a}^*c\mathbf{B}_1-\tilde{a}c\mathbf{B}_1^*)\\
&-\frac{1}{4}(\mathbf{E}_1\cdot\mathbf{J}_{e_1}^*-\mathbf{E}_1^*\cdot\mathbf{J}_{e_1})-\frac{j\omega_{a} }{2} \epsilon_{0} g_{a \gamma \gamma}^2a_0^2 c^2 B_{0}^2.
\end{aligned}
\end{equation}
Then applying the divergence theorem, we arrive at
\begin{equation}
\begin{aligned}
&\oint\operatorname{Re}\left(\mathbf{S}_{DB}\right)\cdot \hat{n}ds=\\
&\int\Big(\frac{j(\omega_1-\omega_a)}{4}\epsilon_0g_{a\gamma\gamma}c\vec{B}_0\cdot(\tilde{a}\mathbf{E}_{1}^*-\tilde{a}^*\mathbf{E}_{1})\\
&+\frac{1}{4}g_{a\gamma\gamma}c\vec{B}_0\cdot(\tilde{a}\mathbf{J}_{e_1}^*+\tilde{a}^*\mathbf{J}_{e_1})-\frac{1}{4}g_{a\gamma\gamma}\vec{J}_{e_0}\cdot(\tilde{a}^*c\mathbf{B}_1+\tilde{a}c\mathbf{B}_1^*)\\
&-\frac{1}{4}(\mathbf{E}_1\cdot\mathbf{J}_{e_1}^*+\mathbf{E}_1^*\cdot\mathbf{J}_{e_1})\Big)~dV,
\end{aligned}
\end{equation}
the same as Eq. (\ref{MinkRe}) in the main text, and
\begin{equation}
\begin{aligned}
&\oint j\operatorname{Im}\left(\mathbf{S}_{DB}\right)\cdot \hat{n}ds=\int\Big(\frac{j\omega_1}{2}\big(\frac{1}{\mu_0}\mathbf{B}_1^* \cdot\mathbf{B}_1-\epsilon_0\mathbf{E}_1\cdot\mathbf{E}_1^*\big)\\
&~+\frac{j(\omega_1+\omega_a)\epsilon_0g_{a\gamma\gamma}}{4}c\vec{B}_0\cdot(\tilde{a}\mathbf{E}_{1}^*+\tilde{a}^*\mathbf{E}_1)~+\\
&\frac{1}{4}g_{a\gamma\gamma}c\vec{B}_0\cdot(\tilde{a}\mathbf{J}_{e_1}^*-\tilde{a}^*\mathbf{J}_{e_1})+\frac{1}{4}g_{a\gamma\gamma}\vec{J}_{e_0}\cdot(\tilde{a}^*c\mathbf{B}_1-\tilde{a}c\mathbf{B}_1^*)\\
&-\frac{1}{4}(\mathbf{E}_1\cdot\mathbf{J}_{e_1}^*-\mathbf{E}_1^*\cdot\mathbf{J}_{e_1}-\frac{j\omega_{a} }{2} \epsilon_{0} g_{a \gamma \gamma}^2a_0^2 c^2 B_{0}^2)\Big)~dV,
\end{aligned}
\end{equation}
the same as Eq. (\ref{MinkIm}) in the main text.

\subsection{Axion modified Abraham Poynting theorem}

To derive Eq. (\ref{AbRe}) and (\ref{AbIm}), we begin with writing the divergence of the real and imaginary parts of $\mathbf{S}_{EH}$ as

\begin{equation}
\begin{aligned}
\nabla\cdot\operatorname{Re}\left(\mathbf{S}_{EH}\right)=\frac{1}{2}\nabla\cdot(\mathbf{S}_{EH}+\mathbf{S}_{EH}^*)\\
\nabla\cdot\operatorname{Im}\left(\mathbf{S}_{EH}\right)=\frac{1}{2}\nabla\cdot(\mathbf{S}_{EH}-\mathbf{S}_{EH}^*)
\end{aligned}
\label{ReImDivEH}
\end{equation}

The next step is to calculate $\nabla\cdot\mathbf{S}_{EH}$ and $\nabla\cdot\mathbf{S}_{EH}^*$,
\begin{equation}
\begin{aligned}
&\nabla \cdot\mathbf{S}_{EH}=\frac{1}{2} \nabla \cdot(\mathbf{E}_{1} \times \frac{1}{\mu_0}\mathbf{B}_1^*) =\\
&\frac{1}{2}\left(\frac{1}{\mu_0}\mathbf{B}_1^* \cdot\nabla \times \mathbf{E}_{1}-\mathbf{E}_{1} \cdot\frac{1}{\mu_0}\nabla \times \mathbf{B}_1^* \right)
\end{aligned}
\label{DivSEH}
\end{equation}
and
\begin{equation}
\begin{aligned}
&\nabla \cdot\mathbf{S}_{EH}^*=\frac{1}{2} \nabla \cdot(\mathbf{E}_{1}^* \times \frac{1}{\mu_0}\mathbf{B}_1) =\\
&\frac{1}{2}\left(\frac{1}{\mu_0}\mathbf{B}_1 \cdot\nabla \times \mathbf{E}_{1}^*-\mathbf{E}_{1}^* \cdot\frac{1}{\mu_0}\nabla \times \mathbf{B}_1 \right).
\end{aligned}
\label{DivSEHIm}
\end{equation}
Considering the Abraham Poynting vector, under a background DC $\it{B}$-$\it{field}$ of, $\vec{B}_0(\vec{r})$, created by an impressed electrical DC current in the magnet coil, $\vec{J}_{e_0}$, in harmonic form, Ampere's law is modified but Faraday's law is not, and may be written as
\begin{equation}
\begin{aligned}
\frac{1}{\mu_0}\nabla \times\mathbf{B}_1&=\mathbf{J}_{e_1}-j\omega_1\epsilon_0\mathbf{E}_{1}+j\omega_ag_{a\gamma\gamma}\epsilon_0c\vec{B}_0\tilde{a}\\
\frac{1}{\mu_0}\nabla \times\mathbf{B}_1^*&=\mathbf{J}_{e_1}^*+j\omega_1\epsilon_0\mathbf{E}_{1}^*-j\omega_ag_{a\gamma\gamma}\epsilon_0c\vec{B}_0\tilde{a}^*\\
\nabla\times\mathbf{E}_{1}&=j\omega_1\mathbf{B}_1\\
\nabla\times\mathbf{E}_{1}^*&=-j\omega_1\mathbf{B}_1^*.
\end{aligned}
\label{AmpFaradayEH}
\end{equation}
Substituting Eqs. (\ref{AmpFaradayEH})  into Eqs. (\ref{DivSEH}) and (\ref{DivSEHIm}) leads to
\begin{equation}
\begin{aligned}
&\nabla \cdot\mathbf{S}_{EH}=\frac{j\omega_1}{2}\left(\frac{1}{\mu_0}\mathbf{B}_1^* \cdot\mathbf{B}_1-\epsilon_0\mathbf{E}_1\cdot\mathbf{E}_1^*\right)\\
&+\frac{j\omega_a}{2}g_{a\gamma\gamma}\epsilon_0\tilde{a}^*c\vec{B}_0\cdot\mathbf{E}_1-\frac{1}{2}\mathbf{E}_1\cdot\mathbf{J}_{e_1}^*
\end{aligned}
\label{divReExEH}
\end{equation}
and
\begin{equation}
\begin{aligned}
&\nabla \cdot\mathbf{S}_{EH}^*=\frac{j\omega_1}{2}\left(\epsilon_0\mathbf{E}_1\cdot\mathbf{E}_1^*-\frac{1}{\mu_0}\mathbf{B}_1^* \cdot\mathbf{B}_1\right)\\
&-\frac{j\omega_a}{2}g_{a\gamma\gamma}\epsilon_0\tilde{a}c\vec{B}_0\cdot\mathbf{E}_1^*-\frac{1}{2}\mathbf{E}_1^*\cdot\mathbf{J}_{e_1}.
\end{aligned}
\label{divImExEH}
\end{equation}
Now by substituting (\ref{divReExEH}) and (\ref{divImExEH}) into (\ref{ReImDivEH}) we obtain
\begin{equation}
\begin{aligned}
\nabla\cdot\operatorname{Re}\left(\mathbf{S}_{EH}\right)&=\frac{j\omega_a}{4}\epsilon_0g_{a\gamma\gamma}c\vec{B}_0\cdot(\tilde{a}^*\mathbf{E}_{1}-\tilde{a}\mathbf{E}_{1}^*)\\
&-\frac{1}{4}(\mathbf{E}_1\cdot\mathbf{J}_{e_1}^*+\mathbf{E}_1^*\cdot\mathbf{J}_{e_1})
\end{aligned}
\end{equation}
and
\begin{equation}
\begin{aligned}
\nabla\cdot j\operatorname{Im}\left(\mathbf{S}_{EH}\right)&=\frac{j\omega_1}{2}\big(\frac{1}{\mu_0}\mathbf{B}_1^* \cdot\mathbf{B}_1-\epsilon_0\mathbf{E}_1\cdot\mathbf{E}_1^*\big)\\
&+\frac{j\omega_a}{4}\epsilon_0g_{a\gamma\gamma}c\vec{B}_0\cdot(\tilde{a}^*\mathbf{E}_{1}+\tilde{a}\mathbf{E}_{1}^*)\\
&-\frac{1}{4}(\mathbf{E}_1\cdot\mathbf{J}_{e_1}^*-\mathbf{E}_1^*\cdot\mathbf{J}_{e_1}).
\end{aligned}
\end{equation}
Then applying the divergence theorem, we arrive at
\begin{equation}
\begin{aligned}
\oint\operatorname{Re}\left(\mathbf{S}_{EH}\right)\cdot \hat{n}ds&=\int\Big(\frac{j\omega_a}{4}\epsilon_0g_{a\gamma\gamma}c\vec{B}_0\cdot(\tilde{a}^*\mathbf{E}_{1}-\tilde{a}\mathbf{E}_{1}^*))\\
&-\frac{1}{4}(\mathbf{E}_1\cdot\mathbf{J}_{e_1}^*+\mathbf{E}_1^*\cdot\mathbf{J}_{e_1})\Big)~dV
\end{aligned}
\end{equation}
and
the same as Eq. (\ref{AbRe}) in the main text, and.
\begin{equation}
\begin{aligned}
\oint j\operatorname{Im}\left(\mathbf{S}_{EH}\right)\cdot \hat{n}ds&=\int\Big(\frac{j\omega_1}{2}\big(\frac{1}{\mu_0}\mathbf{B}_{1}^* \cdot\mathbf{B}_{1}-\epsilon_0\mathbf{E}_1\cdot\mathbf{E}_1^*\big)\\
&+\frac{j\omega_a}{4}\epsilon_0g_{a\gamma\gamma}c\vec{B}_0\cdot(\tilde{a}^*\mathbf{E}_{1}+\tilde{a}\mathbf{E}_{1}^*)\\
&-\frac{1}{4}(\mathbf{E}_1\cdot\mathbf{J}_{e_1}^*-\mathbf{E}_1^*\cdot\mathbf{J}_{e_1}))\Big)~dV,
\end{aligned}
\end{equation}
the same as Eq. (\ref{AbIm}) in the main text.

\section{APPENDIX B: CONSIDERATION OF THE REACTIVE POWER FLOW IN A RESONANT HALOSCOPE}

\begin{figure}[t!]
\includegraphics[width=0.48\textwidth]{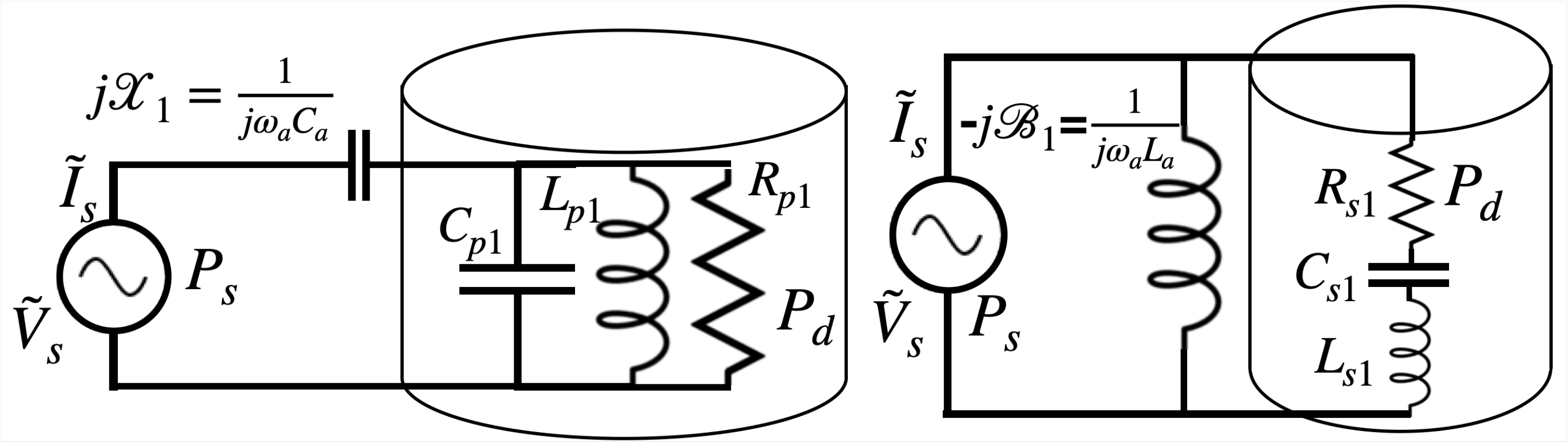}
\caption{Left, equivalent parallel $LCR$ circuit representation of an axion coupling to a resonant cavity haloscope. Right, the equivalent series $LCR$ circuit representation.}
\label{CavHaloCct}
\end{figure}

In this appendix we consider the impact of the reactive part of the Poynting vector on a resonant cavity axion haloscope. In general reactive coupling of power into a resonant cavity may be calculated by implementing Foster's reactance theorem \cite{Foster1924,Dicke1987,Beringer1987}. Foster showed that a lossless circuit network made of resonances and antiresonances could be represented as a combination of inductors and capacitors\cite{Foster1924} and following this Beringer and Dicke applied the theorem to high-$Q$ microwave cavities\cite{Dicke1987,Beringer1987}, allowing the calculation of the effect of reactive coupling to a cavity, based on the complex Poynting theorem \cite{Montgomery87}. In general, it has been shown that the reactive coupling network into a high-$Q$ resonance may be represented by either an impedance, $\mathcal{X}_1$ in series with a parallel LC circuit (of elements $L_{p1}$, $C_{pi}$, and $R_{p1}$), or an admittance, $\mathcal{B}_1$, in parallel with a series LC circuit (of elements $L_{s1}$, $C_{s1}$, and $R_{s1}$). Applying this technique to axion-photon conversion in a resonant haloscope, leads to the following equivalent circuit shown in Fig. \ref{CavHaloCct}.

Applying Foster's reactance theorem to the cavity resonator allows $\mathcal{X}_1$ and $\mathcal{B}_1$ to be calculated from the following equations \cite{Beringer1987}:
\begin{equation}
\begin{aligned}
j\mathcal{X}_1=j\omega_1\frac{4\left(U_{B1}-U_{E1}\right)}{\tilde{I} \tilde{I}^{*}}~\text{and}~j\mathcal{B}_1=j\omega_1\frac{4\left(U_{E1}-U_{B1}\right)}{\tilde{\mathcal{V}} \tilde{\mathcal{V}} ^{*}},
\end{aligned}
\label{jX}
\end{equation}
where
\begin{equation}
\begin{aligned}
U_{B1}=\frac{1}{4\mu_0}\int\mathbf{B}_{1}^* \cdot\mathbf{B}_{1}~dV~\text{and}~U_{E1}=\frac{\epsilon_0}{4}\int\mathbf{E}_1\cdot\mathbf{E}_1^*~dV.
\end{aligned}
\label{jX2}
\end{equation}
Following this procedure, the axion-photon coupling input impedances for both the axion modified Abraham Poynting Vector and the axion modified Minkowski Poynting Vector may be calculated, and this is undertaken in the following sections.

\subsubsection{Abraham Poynting theorem}

To calculate the parameters for the parallel $LCR$ circuit shown in Fig.\ref{CavHaloCct}, Eqs.(\ref{jX}) and (\ref{AbIm}) are combined, and given that the real part of $\mathbf{E}_1$ is out of phase with any conduction currents in the volume, and zero at the cavity boundary, then the series impedance becomes,
\begin{equation}
\begin{aligned}
j\mathcal{X}_1=-j\frac{\omega_a}{2}\frac{\int\epsilon_0g_{a\gamma\gamma}c\vec{B}_0\cdot(\tilde{a}^*+\tilde{a})Re(\mathbf{E}_{1})~dV}{\tilde{I} \tilde{I}^{*}}.
\end{aligned}
\label{jXa}
\end{equation}
Then given that the energy in a LC resonator is given by $U_1=\frac{1}{2}\tilde{I} \tilde{I}^{*}L_{p1}$, where $\omega_1^2=\frac{1}{L_{p1}C_{p1}}$, and using the result from Eq. (\ref{U1}), Eq. (\ref{jXa}) becomes
\begin{equation}
\begin{aligned}
j\mathcal{X}_1=\frac{\omega_a}{j\omega_1^2C_{p1}}\frac{\frac{a_0\epsilon_0g_{a\gamma\gamma}c}{2}\int\vec{B}_0\cdot Re(\mathbf{E}_{1})~dV}{U_1}=\frac{\omega_a}{j\omega_1^2C_{p1}},
\end{aligned}
\label{jXaa}
\end{equation}
which is equivalent to a capacitance of $C_a=C_{p1}\frac{\omega_1^2}{\omega_a^2}$. Thus, the input impedance, $Z_p(\omega_a)$, of the parallel circuit representation can be written in normalized form as
\begin{equation}
\begin{aligned}
\frac{Z_p(\omega_a)}{R_{p1}}=\frac{\omega_a}{j\omega_1Q_1}+\frac{1}{1+j Q_1\left(\frac{\omega_a}{\omega_{1}}-\frac{\omega_{1}}{\omega_a}\right)},
\end{aligned}
\label{Zw}
\end{equation}
where $Q_1=\omega_1R_{p1}C_{p1}$. Defining the detuning as $\delta\omega=\omega_a-\omega_1$, where $\delta\omega\ll\omega_1$ and $\delta_a=\frac{\delta\omega}{\omega_1}$, then
\begin{equation}
\begin{aligned}
\frac{Z_p(\delta_a)}{R_{p1}}\approx\frac{1-j2Q_1(\delta_a+\frac{1}{2Q_1^2})}{1+4Q_1^2\delta_a^2},
\end{aligned}
\label{Zww}
\end{equation}
Setting the imaginary part to zero allows the calculation of the frequency shift of the resonant mode due to the axion coupling, which gives $\frac{\delta\omega_1}{\omega_1}\sim-\frac{1}{2Q_1^2}$ a very small frequency shift, which to first order does not affect the sensitivity of the axion haloscope and is the same order as a frequency shift due to dissipation. Ignoring this term gives the usual complex response of a resonant LCR circuit.

To calculate the parameters for the series $LCR$ circuit shown in Fig. \ref{CavHaloCct}, we can use a similar procedure given $U_1=\frac{1}{2}\tilde{\mathcal{V}} \tilde{\mathcal{V}}^{*}C_{s1}$, where, $\omega_1=\frac{1}{L_{s1}C_{s1}}$, along with Eqs. (\ref{jX}) and (\ref{AbIm}) to show that the parallel input admittance is given by,
\begin{equation}
\begin{aligned}
\text{-}j\mathcal{B}_1=\frac{\omega_a}{j\omega_1^2L_{s1}}\frac{\frac{a_0\epsilon_0g_{a\gamma\gamma}c}{2}\int\vec{B}_0\cdot Re(\mathbf{E}_{1})~dV}{U_1}=\frac{\omega_a}{j\omega_1^2L_{s1}},
\end{aligned}
\label{jBaa}
\end{equation}
which is equivalent to an inductance of $L_a=L_{s1}\frac{\omega_1^2}{\omega_a^2}$. Thus, the input admittance, $Y_s(\omega_a)$, of the series circuit representation may be written in normalized form as,
\begin{equation}
\begin{aligned}
R_{s1}Y_s(\omega_a)=\frac{\omega_a}{j\omega_1Q_1}+\frac{1}{1+j Q_1\left(\frac{\omega_a}{\omega_{1}}-\frac{\omega_{1}}{\omega_a}\right)},
\end{aligned}
\label{Yww}
\end{equation}
where $Q_1=\frac{\omega_1L_{s1}}{R_{s1}}$, so
\begin{equation}
\begin{aligned}
R_{s1}Y_s(\omega_a)\approx\frac{1-j2Q_1(\delta_a+\frac{1}{2Q_1^2})}{1+4Q_1^2\delta_a^2},
\end{aligned}
\label{Yww}
\end{equation}
which completes the dual representation of the resonant axion haloscope as either a parallel $LCR$ in series with a capacitive coupling element or a series $LCR$ circuit in parallel with an inductive coupling element.

\subsubsection{Minkowski Poynting theorem}

The reactive part of the Minkowski Poynting vector as written in Eq. (\ref{MinkIm}) has extra terms compared to the Abraham Poynting vector, and by following a similar process, the equivalent equation for the series impedance can be calculated to be,
\begin{equation}
\begin{aligned}
j\mathcal{X}_{1M}&=\frac{\omega_a+\omega_1}{{j\omega_1^2C_{p1}}}\frac{\frac{\epsilon_0g_{a\gamma\gamma}a_0c}{2}\int\vec{B}_0\cdot Re(\mathbf{E}_{1})~dV}{U_1}\\
&+\frac{1}{4{\omega_1^2C_{p1}}}\frac{g_{a\gamma\gamma}c\int\vec{B}_0\cdot(\tilde{a}^*\mathbf{J}_{e_1}-\tilde{a}\mathbf{J}_{e_1}^*)dV}{U_1} \\
&-\frac{\omega_a}{{j\omega_1^2C_{p1}}}\frac{\epsilon_0g_{a\gamma\gamma}^2a_0^2c^2B_0^2V_1}{U_1}.
\end{aligned}
\label{jXaM}
\end{equation}
The second term is nonzero due to lossless inductive currents at the cavity surface, $\kappa_{e_1}$, which are in phase with the magnetic field, $\mathbf{B}_1$, and related by $\kappa_{e 1}=\frac{1}{\mu_{0}} \hat{n} \times \mathbf{B}_{1}$, where $\hat{n}$ is the normal to the cavity surface, and because the surface current and magnetic field are in imaginary phase, then $\kappa_{e_1}^*=\text{-}\kappa_{e_1}$. Note $\mathbf{J}_{e_1}$=0, over the volume, unless there is loss in the volume, which contributes to the real part of the Poynting vector, not the reactive part. Next, by implementing the identity, $\oint d \vec{s}\times\mathbf{B}_1=\int \nabla \times \mathbf{B}_1~dV$, and from Eq. (\ref{PhasorAmp}), to first order we may substitute the following, $\nabla \times \mathbf{B}_1\rightarrow-j\omega_1\epsilon_0\mathbf{E}_{1}$ (ignoring terms second order in $g_{a\gamma\gamma}$). Then it is straightforward to show (given $\vec{ds}=ds\hat{n}$)
\begin{equation}
\begin{aligned}
&\vec{B}_0\cdot\int(\tilde{a}^*\mathbf{J}_{e_1}-\tilde{a}\mathbf{J}_{e_1}^*)~dV=2a_0\vec{B}_0\cdot\oint\mathbf{\kappa}_{1}~ds\\
&=2a_0\vec{B}_0\cdot\oint\vec{ds}\times\mathbf{B}_{1}=-2ja_0\omega_1\int\vec{B}_0\cdot Re(\mathbf{E}_{1})~dV,
\end{aligned}
\label{jXaMM}
\end{equation}
for the resonant cavity haloscope. Therefore substituting Eq. (\ref{jXaMM}) into  (\ref{jXaM}), the series impedance becomes,
\begin{equation}
\begin{aligned}
j\mathcal{X}_{1M}&=\frac{\omega_a+2\omega_1-\frac{2\omega_a}{C_1}}{{j\omega_1^2C_{p1}}}
\end{aligned}
\label{jXaM2}
\end{equation}
where the $C_1$ is the form factor from (\ref{FormFac}).
So the effective input capacitance represented in Fig.\ref{CavHaloCct} becomes $C_a=C_{p1}\frac{\omega_1}{\omega_a(2-\frac{\omega_a}{\omega_1}(\frac{2}{C_1}-1))}$, which is of the same order as $C_a$ for the Abraham equivalent circuit when $\omega_a\sim\omega_1$. The normalized input impedance to first order in $\delta_a$ can thus be written as
\begin{equation}
\begin{aligned}
\frac{Z_p(\delta_a)}{R_{p1}}\approx\frac{1-j2Q_1(\delta_a+\frac{3-\frac{2}{C_1}}{2Q_1^2})}{1+4Q_1^2\delta_a^2}
\end{aligned}
\label{ZwwM}
\end{equation}
Setting the imaginary part to zero allows the calculation of the frequency shift of the resonant mode due to the axion coupling, which gives $\frac{\delta\omega_1}{\omega_1}\sim\text{-}\frac{3-\frac{2}{C_1}}{2Q_1^2}$ a very small frequency shift on the same order of the Abraham Poynting vector prediction. A precision frequency measurement of the axion interacting with a microwave cavity haloscope would be needed to determine this frequency shift.

A similar calculation can be undertaken for the effective parallel inductance for the series $LCR$ circuit representation, the end result is an inductance of $L_a=L_{s1}\frac{\omega_1}{\omega_a(2-\frac{\omega_a}{\omega_1}(\frac{2}{C_1}-1))}$ leading to similar conclusions and a normalized input admittance of,
\begin{equation}
\begin{aligned}
R_{s1}Y_s(\omega_a)\approx\frac{1-j2Q_1(\delta_a+\frac{3-\frac{2}{C_1}}{2Q_1^2})}{1+4Q_1^2\delta_a^2},
\end{aligned}
\label{YwwM}
\end{equation}
which completes our analysis. \\

\end{document}